\documentclass[english,12pt]{article}
\usepackage[T1]{fontenc}
\usepackage[latin9]{inputenc}
\usepackage[a4paper]{geometry}
\usepackage{color}
\definecolor{note_fontcolor}{rgb}{0.800781, 0.800781, 0.800781}
\usepackage{pifont}
\usepackage{array}
\usepackage{tabularray}
\usepackage{verbatim}
\usepackage{rotfloat}
\usepackage{multirow}
\usepackage{url}
\usepackage{booktabs}
\usepackage{mathtools}
\usepackage{lscape}
\usepackage{algpseudocode}
\usepackage{amstext}
\usepackage{amsthm}
\usepackage{graphicx}
\usepackage{longtable}
\usepackage{setspace}
\usepackage{float}
\usepackage{comment}
\usepackage{hhline}
\usepackage{caption}
\usepackage{adjustbox}
\usepackage{mathalpha}
\usepackage{algorithm}

\usepackage{natbib}
\usepackage{amsfonts}
\usepackage{mathrsfs}
\usepackage{amsmath}
\usepackage{subfigure}
\usepackage{hyperref}
\usepackage{booktabs}
\usepackage{multirow}
\usepackage{siunitx}
\definecolor{revcol}{rgb}{0,0,1}  
\usepackage{pdfpages}

\begin{document}
\title{Circular interval-valued regression with application to wind direction: A cylindrical approach }
\author{Thasmika Mohan$^{1}$, Najmeh Nakhaei Rad$^{1\footnote{Corresponding Author: najmeh.nakhaeirad@up.ac.za}}$, Sphiwe Skhosana$^{1}$ \\
$^{1}$ {\small Department of Statistics, University of Pretoria, Pretoria 0002, South Africa.}}
\date{}
	\maketitle
	\abstract{Many real-world datasets are inherently interval-valued, yet standard statistical methods are typically designed for point observations. Furthermore, intervals that are circular in nature may naturally exist. In this paper, we introduce a cylindrical regression framework for interval-valued circular data, addressing the joint challenges of the periodical nature and measurement uncertainty within the observed circular intervals. The proposed model accommodates a circular interval-valued response variable and both circular and linear covariates using a centre-range approach. To estimate the model, a maximum likelihood approach is proposed. The performance of the proposed estimation procedure is demonstrated through a simulation study. Application to two real-world datasets from a wind farm in South Africa and a weather station in Brazil further substantiates the model's practical value.\\\\
	}	
	{{\bf Keywords}: Circular interval-valued data; circular-linear regression; Maximum likelihood estimation; Cylindrical distributions, Centre-range method}
	
	\section{Introduction}
    In many modern applications, observations are not precise values but intervals that reflect uncertainty and variability \citep{billard2003symbolic}. Such data are seen across various domains, and common examples include: monitoring of daily temperature fluctuations, representation of hourly exchange rate volatility between the Euro and the US Dollar, recording of spatiotemporal pollution levels across diverse geographical locations, and the quantification of mineral concentration ranges within food items \citep{leski2016fuzzy}. While these examples represent observed intervals on the real line, similar interval-valued data can be seen on the circle. 
   Observations recorded on the circle, often expressed as directions or angles, are located on the unit circle, where the starting and ending points coincide (e.g., $0^\circ$ and $360^\circ$). Data of this nature are periodic and can be found across many scientific disciplines, such as meteorology, biology, geology, and engineering \citep{mardia2000directional}, for example, wind direction data  \citep{rad2022enhancing,nakhaei2026predicting,ridder2025power}, ocean-current directions \citep{mohan2024poly}, animal movement bearings
\citep{ramkilawon2023sine,bekker2022generalized,gascoyne2024modelling}, molecular angles
\citep{arashi2021mobius,rad2025bayesian} and the timing of periodic events
\citep{skhosana2026model}.
   
 Currently, classical circular regression frameworks are formulated under the assumption of point-wise observations and therefore cannot directly accommodate interval-valued data. This limitation poses a significant methodological challenge, as each interval-valued observation has a degree of uncertainty that should be preserved in the modelling process.

Furthermore, linear techniques do not respect the circular topology and may introduce discontinuities at the circular cut.  
The simplest aggregation method replaces each interval $[a_i, b_i]$ with its midpoint $m_i = (a_i + b_i)/2$ \citep{billard2003symbolic}. More advanced approaches incorporate both the midpoint and the range $r_i = (b_i - a_i)/2$ as explanatory components, enabling the joint modelling of the central tendency and dispersion of the intervals \citep{neto2008centre}. Other formulations treat the lower and upper bounds as correlated variables in a bivariate framework \citep{billard2007symbolic} or use symbolic and stochastic representations to preserve within-interval variability \citep{arroyo2009interval}. These methods are also applied to the analysis of circular interval-valued data \citep{mixcirc2026}. This motivates the need to develop circular interval-valued regression models. 

In this paper, we introduce a circular centre-range (CCR) regression model to express the relationship between a circular interval-valued response variable and several interval-valued independent variables, including both linear and circular types. A maximum likelihood estimation (MLE) approach is proposed to estimate the parameters of the model. A simulation study is conducted to evaluate the performance of the proposed estimation procedure using different scenarios. The practical utility of the CCR regression model is motivated by applications to interval-valued data from the Nouport wind farm in South Africa and a weather station in Brazil.

The rest of this paper is organised as follows: Section~\ref{sec:regression} provides an overview of existing circular-linear regression models, followed by Section \ref{sec:methodology} where the proposed CCR model and its estimation procedure are discussed. Section~\ref{sec:simulation} describes a simulation study conducted to evaluate the performance of the proposed estimation approach. Finally,  Section~\ref{sec:application} illustrates the practical usefulness of the proposed approach using data from South Africa and from Brazil.

\section{Background on circular-linear regression}\label{sec:regression} 
	Circular regression encompasses a family of statistical models suitable to explain relationships in which at least one variable is circular. 
	The classification of circular regression models depend on the nature of the variables involved: models with both circular response and predictors are termed circular-circular, whereas those with a circular response and linear predictors are referred to as circular-linear, and vice versa for linear-circular models \citep{mardia2000directional}. Some examples of using circular regression models in practice are as follows: in meteorology, wind direction has been modelled as a function of pollutant concentration and wind speed 
	\citep{fisher1992regression, sengupta2006asymmetric, mohammad2021review}; in oceanography, the wave direction has been modelled in relation to ocean currents and wind speed \citep{gould1969regression}; in medicine, circular regression has been used to analyse vector cardiogram orientations as a function of heartbeat duration 
	\citep{ mardia2000directional}; in neuroscience and motor control research, hand movement direction has been modelled as a function of reaction time and force magnitude \citep{berens2009circular}; in ecology, circular regression models have been applied to model animal movement and orientation behaviours 
	\citep{batschelet1981circular}; in meteorology, air quality has been modelled using other environmental factors \citep{kim2017multivariate}; in biology the circadian rhythms of gene expression cycles is modelled using genetic factors \cite{hughes2010jtk_cycle}. Further applications of circular regression include geological studies of fault and fracture orientations \citep{mardia2000directional} 
	and an astronomical analysis of angular measurements, such as celestial coordinates \citep{mardia2000directional}.

    The foundational theory of circular-linear regression was introduced by \cite{gould1969regression}, who developed a regression model to predict a circular response variable based on linear predictors. The circular response variable was assumed to follow a von Mises distribution, analogous to the Gaussian distribution on the real line. Building on this, \cite{johnson1978some} showed that the maximum likelihood function from Gould's model has infinitely many equally high peaks, which can lead to ambiguously defined MLEs. Hence, these authors addressed this with a single linear predictor and demonstrated with an application in meteorology. Subsequently, \cite{fisher1992regression} proposed three extensions of Gould's model to incorporate a concentration parameter for the response variable whilst allowing for multiple linear predictors through a link function that transforms values from the real line onto the unit circle. Together, the models of \cite{gould1969regression}, \cite{johnson1978some}, and \cite{fisher1992regression} laid the parametric foundation for circular-linear regression. 
	\cite{lund1999least} presented a circular regression model with a combination of circular and linear predictors and a circular response variable. However, the link function limits how far the explanatory variables can rotate the mean
    direction, and the fitted model depends on the arbitrary origin of each linear covariate
    \citep{fisher1992regression}. Consequently, linear covariates are often rescaled
    before estimation \citep{cremers2020cylindrical}, whereas circular covariates enter through
    their cosine and sine. In later years, \cite{george2006semiparametric} presented a semi-parametric Bayesian model for circular-linear regression. Further extensions of \cite{gould1969regression} and \cite{johnson1978some} can be seen in \cite{fernandez2007models}, where the author developed a new model based on non-negative trigonometric sums. Thereafter, \cite{sengupta2006asymmetric} introduced an asymmetric circular-linear regression model for environmental data. \cite{kim2015inverse} developed an inverse circular-linear regression approach. 
	For a comprehensive review of circular-linear regression approaches, see \cite{mohammad2021review}. \cite{alonso2023analyzing} considered the use of non-parametric circular regression for multi-modal data. Recently, \cite{mixcirc2026} proposed a mixture of circular-linear regression models to capture the multimodality in the circular response variable with an application to South African wind data.

    To the best of our knowledge, there are currently no regression models for circular interval-valued data. To fill this gap in the literature, we introduce the CCR regression model in the next section.  

\section{Methodology}\label{sec:methodology}
In this section, we propose a centre--range framework that captures both the circular
topology and the interval structure of the observations and present a maximum likelihood estimating approach for the proposed model. Drawing on the centre--range approach of \citet{neto2008centre}, we develop a regression framework for circular interval-valued responses. The framework builds on the circular--linear regression model of \citet{fisher1992regression} and considers three alternative specifications for the joint distribution of the centre and range, based on the cylindrical distributions of \citet{mohan2026interval}.

Let $\Theta=[\Theta^L,\Theta^U]$ denote a circular interval-valued response variable, with
realisations $\theta_i=[\theta_i^L,\theta_i^U]$, $i=1,\dots,n$, where
$\theta_i^L,\theta_i^U\in[-\pi,\pi)$ are the lower and upper bounds of the $i$th observed
interval. Each circular interval is defined as the anticlockwise arc from its lower endpoint \(\theta_i^L\) to its upper endpoint \(\theta_i^U\), with both endpoints represented in \([-\pi,\pi)\).  Using a centre--range representation of the intervals,  let each response interval be represented as a point $(C_i,Q_i)$ on the cylinder
$[-\pi,\pi)\times(0,1)$, so that $c_i$ and $q_i$ are their observed values. The centre of an interval, $C_i$ is defined as
\begin{equation}\label{eq:centre}
  C_i =
  \begin{cases}
    \mathrm{mod}_{2\pi}\!\left(\dfrac{\theta_i^L+\theta_i^U}{2}+\pi\right)-\pi,
      & \theta_i^U \ge \theta_i^L,\\[2ex]
    \mathrm{mod}_{2\pi}\!\left(\dfrac{\theta_i^L+\theta_i^U}{2}\right)-\pi,
      & \theta_i^U < \theta_i^L,
  \end{cases}
\end{equation}
and $Q_i$ is the bounded range of an interval and is defined as
\begin{equation}\label{eq:range}
  Q_i = \frac{\mathrm{mod}_{2\pi}\big(\theta_i^U-\theta_i^L\big)}{2\pi}.
\end{equation}
The bounds can be recovered from the centre and range as
\begin{equation*}
  \theta_i^L = \mathrm{mod}_{2\pi}\big(C_i-\pi Q_i+\pi\big)-\pi, \qquad
  \theta_i^U = \mathrm{mod}_{2\pi}\big(C_i+\pi Q_i+\pi\big)-\pi.
\end{equation*}

\subsection{Cylindrical distributions for the centre and range}
In \citet{mohan2026interval}, for time series analysis, the centre and range of  circular intervals were modelled jointly
on the cylinder, using a von Mises distribution for the centre and one of three bounded
distributions for the range, namely the truncated Weibull, reparameterised beta \citep{ferrari2004beta} and
Kumaraswamy \citep{kumaraswamy1980} distributions, respectively. In this section, we adopt these cylindrical distributions within a
regression framework instead. Assuming the centre $C_i$ follows a von Mises distribution,  the density is
\begin{equation}
f_{C\mid Q}(c_i\mid q_i)
=
\frac{\exp\!\left\{\upsilon(q_i)\cos\!\left(c_i-\mu_i\right)\right\}}
{2\pi I_0\!\left(\upsilon(q_i)\right)},
\label{eq:vm}
\end{equation}
where the mean direction is $\mu_i$, the concentration parameter is $\upsilon(q_i)>0$, and $I_0(\cdot)$ is the modified Bessel function of the first kind of order zero. We consider the same three distributions for the bounded range as in
\citet{mohan2026interval}:

\begin{itemize}
    \item The truncated Weibull distribution with density
\[
f_Q(q_i)
=
\frac{
\dfrac{r}{\lambda_i}
\left(\dfrac{q_i}{\lambda_i}\right)^{r-1}
\exp\!\left[-\left(\dfrac{q_i}{\lambda_i}\right)^r\right]
}{
1-\exp\!\left[-\lambda_i^{-r}\right]
},
\]
where $r>0$ is the shape parameter and $\lambda_i=\exp(\eta_i)>0$ is the scale parameter. \item The reparameterised beta  distribution \citep{ferrari2004beta} has the following density
\[
f_Q(q_i)
=
\frac{\Gamma(\tau)}{\Gamma(m_i\tau)\,\Gamma\!\left((1-m_i)\tau\right)}
q_i^{m_i\tau-1}(1-q_i)^{(1-m_i)\tau-1},
\] where the mean is $m_i\in(0,1)$ and precision
$\tau>0$. 
\item The Kumaraswamy distribution \citep{kumaraswamy1980} with density
\[
f_Q(q_i)
=
a\,b_i\,q_i^{a-1}\left(1-q_i^a\right)^{b_i-1},
\]
where $a>0$ and $b_i=\exp(\eta_i)>0$ are shape parameters. 
\end{itemize}
For each model, the joint cylindrical density is obtained as \citep{harvey2023regime}
\begin{equation}
\begin{aligned}
f(c_i,q_i)
&=
f_Q(q_i)\,f_{C\mid Q}(c_i\mid q_i)\\
&=
f_Q(q_i)\,
\frac{
\exp\!\left\{\upsilon(q_i)\cos\!\left(c_i-\mu_i\right)\right\}
}{
2\pi I_0\!\left(\upsilon(q_i)\right)
},
\end{aligned}
\label{eq:cr-density}
\end{equation}
As in \citet{harvey2023regime}, the concentration of the centre depends on the range through $\upsilon(q_i)=\exp(\zeta-\delta q_i),~\zeta,\delta\in\mathbb R$ and the log-likelihood function is
\begin{equation}
\ell_i
={}
\log f_Q(q_i)
+\upsilon_i(q_i)
\cos\!\left(c_i-\mu_i\right)
-\log(2\pi)
-\log I_0\!\left(\upsilon_i(q_i)\right).
\label{eq:cr-loglik}
\end{equation}

The parameter \(\delta\) governs the dependence between the centre and range of the circular intervals through the concentration parameter of the conditional von Mises distribution. Specifically, \(\delta>0\) implies that increasing interval width is associated with decreasing concentration of the centre around its conditional mean direction, whereas \(\delta<0\) implies increasing concentration. When \(\delta=0\), the concentration parameter no longer depends on the interval range, and the centre and range are conditionally independent given the covariates under the proposed model.
\subsection{Centre-range regression model for circular data}
In this section, we present the proposed circular centre--range (CCR) regression model. Assume the linear interval-valued covariates are defined as
$\mathbf{Z}=(Z_1,\dots,Z_l)^{\top}$, with realisations $z_{ik}=[z_{ik}^L,z_{ik}^U]$,
$k=1,\dots,l$, and the circular interval-valued covariates are defined as
$\mathbf{\Phi}=(\Phi_1,\dots,\Phi_s)^{\top}$, with realisations
$\phi_{ij}=[\phi_{ij}^L,\phi_{ij}^U]$, $j=1,\dots,s$. Each linear covariate is represented by its
linear centre, $z_{ik}^C=\frac{z_{ik}^L+z_{ik}^U}{2}$ and linear range $z_{ik}^R=z_{ik}^U-z_{ik}^L$, respectively.
Similarly, each circular covariate is represented by its circular centre, $\phi_{ij}^C$ and bounded range,
$\phi_{ij}^R$ which can be obtained by using \eqref{eq:centre} and \eqref{eq:range}. Hence, the centre representations of the circular and linear
covariates are given by
\begin{align*}
\mathbf{x}_i^C &= \big(z_{i1}^C,\dots,z_{il}^C,\
  \cos\phi_{i1}^C,\ \sin\phi_{i1}^C,\dots,\cos\phi_{is}^C,\ \sin\phi_{is}^C\big)^{\top},
\end{align*}
with corresponding coefficients for the centres
\begin{align*}
\boldsymbol{\beta}^C &= \big(\beta_1^C,\dots,\beta_l^C,\
  \gamma_{11}^C,\ \gamma_{12}^C,\dots,\gamma_{s1}^C,\ \gamma_{s2}^C\big)^{\top},
\end{align*}
where $\mathbf{x}_i^C$ and $\boldsymbol{\beta}^C$ are $(l+2s)\times 1$ vectors, respectively. Similarly, the covariates for the ranges are given by
\begin{align*}
\mathbf{x}_i^R &= \big(z_{i1}^R,\dots,z_{il}^R,\ \phi_{i1}^R,\dots,\phi_{is}^R\big)^{\top},
\end{align*}
with corresponding coefficients,
\begin{align*}
\boldsymbol{\beta}^R &= \big(\beta_1^R,\dots,\beta_l^R,\ \gamma_1^R,\dots,\gamma_s^R\big)^{\top},
\end{align*}
where $\mathbf{x}_i^R$ and $\boldsymbol{\beta}^R$ are $(l+s)\times 1$ vectors, respectively.

Therefore, the CCR model is defined as 
\begin{equation}
\begin{pmatrix}\mu_i\\[2pt] \eta_i\end{pmatrix}
=
\begin{pmatrix}
\mathcal{M}^C+2\arctan\big({\boldsymbol{\beta}^{C}}^{\top}\mathbf{x}_i^C\big)\\[2pt]
\mathcal{M}^R+{\boldsymbol{\beta}^{R}}^{\top}\mathbf{x}_i^R
\end{pmatrix}, \label{model}
\end{equation}
where $\mathcal{M}^C\in[-\pi,\pi), \mathcal{M}^R \in \mathbb{R} $ are the intercept terms, $\mu_i$ is the mean direction of the centre in \eqref{eq:vm} and $\eta_i$ differs for each range distribution. Hence $\eta_i$ is specified below for each distribution:
\begin{itemize}
\item Truncated Weibull: The scale parameter is expressed via a log-link as
\[ \eta_i=
\log\lambda_i=\mathcal{M}^R+\boldsymbol{\beta}^{R\top}\mathbf{x}_i^R,
\qquad
\lambda_i=\exp(\eta_i)>0 .
\]
The conditional expected range is
\[
\mathbb{E}\big(Q_i\mid\mathbf{x}_i^R\big)=\lambda_i\,
\frac{\gamma\!\left(1+\frac{1}{r},\,\lambda_i^{-r}\right)}{1-\exp\!\left(-\lambda_i^{-r}\right)},
\]
where $\gamma(\cdot,\cdot)$ is the lower incomplete gamma function.
This distribution coupled with the von-Mises distribution in \eqref{eq:cr-density} is called the CCR Truncated Weibull von-Mises (CCR-TWvM) model.
\item Reparameterised beta:  The mean parameter is expressed via a logit link,
\[ \eta_i=
\operatorname{logit}(m_i)=\log\frac{m_i}{1-m_i}=\mathcal{M}^R+\boldsymbol{\beta}^{R\top}\mathbf{x}_i^R,
\qquad
m_i=\frac{1}{1+\exp(-\eta_i)}\in(0,1).
\]
The conditional expected range is
\[
\mathbb{E}\big(Q_i\mid\mathbf{x}_i^R\big)=m_i .
\]
Together with the von-Mises distribution in \eqref{eq:cr-density}, this model is called the CCR beta von-Mises (CCR-BvM) model.
\item Kumaraswamy: The second shape parameter is expressed via a log-link function:
\[
\eta_i=\log b_i=\mathcal{M}^R+\boldsymbol{\beta}^{R\top}\mathbf{x}_i^R,
\qquad
b_i=\exp(\eta_i)>0 .
\]
The conditional expected range is
\[
\mathbb{E}\big(Q_i\mid\mathbf{x}_i^R\big)=b_i\,B\!\left(1+\frac{1}{a},\,b_i\right),
\]
where $B(\cdot,\cdot)$ is the beta function.
This distribution coupled with the von-Mises distribution in \eqref{eq:cr-density} is called the CCR Kumaraswamy von-Mises (CCR-KvM) model.
\end{itemize}

\subsection{Maximum likelihood estimation}

For each model, let the parameters be
$
\boldsymbol{\psi}=\begin{pmatrix}
\boldsymbol{\psi}_C\\
\boldsymbol{\psi}_Q
\end{pmatrix},
$
where the centre parameter vector is
$
\boldsymbol{\psi}_C
=
(\mathcal{M}^C,{\boldsymbol{\beta}^{C}}',\zeta,\delta)'
$
and the range parameter vector depends on the chosen distribution,
\begin{align*}
\text{CCR-TWvM:}\qquad
\boldsymbol{\psi}_Q
&=
(\mathcal{M}^R,{\boldsymbol{\beta}^{R}}',\log r)',\\
\text{CCR-BvM:}\qquad
\boldsymbol{\psi}_Q
&=
(\mathcal{M}^R,{\boldsymbol{\beta}^{R}}',\log\tau)',\\
\text{CCR-KvM:}\qquad
\boldsymbol{\psi}_Q
&=
(\mathcal{M}^R,{\boldsymbol{\beta}^{R}}',\log a)'.
\end{align*}
For $n$ observations, the conditional
log-likelihood is
\begin{align}
\ell(\boldsymbol{\psi};\mathbf c,\mathbf q
     \mid\mathbf X^C,\mathbf X^R)
&=
\ell_Q(\boldsymbol{\psi}_Q;\mathbf q\mid\mathbf X^R)
+
\ell_{C\mid Q}(\boldsymbol{\psi}_C;\mathbf c
               \mid\mathbf q,\mathbf X^C)
\label{eq:loglik}\\
&=
\sum_{i=1}^{n}
\log f_Q(q_i\mid\mathbf x_i^R;\boldsymbol{\psi}_Q)
\notag\\
&\quad+
\sum_{i=1}^{n}
\left[
\upsilon(q_i)\cos(c_i-\mu_i)
-\log(2\pi)
-\log I_0\!\left(\upsilon(q_i)\right)
\right],
\notag
\end{align}

and the log-likelihood of the range depends on the chosen distribution 

\begin{itemize}
\item {CCR-TWvM:}
\[
\ell_Q(\boldsymbol{\psi}_Q)=\sum_{i=1}^{n}\Big[\log r-r\eta_i+(r-1)\log q_i
-\Big(\frac{q_i}{\lambda_i}\Big)^{r}-\log\big\{1-\exp(-\lambda_i^{-r})\big\}\Big],
\]
\item {CCR-BvM:}
\[
\begin{aligned}
\ell_Q(\boldsymbol{\psi}_Q)=\sum_{i=1}^{n}\Big[&\log\Gamma(\tau)-\log\Gamma(m_i\tau)
-\log\Gamma\{(1-m_i)\tau\}\\
&+(m_i\tau-1)\log q_i+\{(1-m_i)\tau-1\}\log(1-q_i)\Big],
\end{aligned}
\]
\item {CCR-KvM:}
\[
\ell_Q(\boldsymbol{\psi}_Q)=\sum_{i=1}^{n}\Big[\log a+\eta_i+(a-1)\log q_i
+(b_i-1)\log\big(1-q_i^{a}\big)\Big].
\]
\end{itemize}
The maximum
likelihood estimate $\hat{\boldsymbol{\psi}}=\arg\max_{\boldsymbol{\psi}}\ell(\boldsymbol{\psi})$ has no closed-form expression
and is obtained numerically using a quasi-Newton algorithm. As noted by
\citet{fisher1992regression}, the log-likelihood of the centre equation can have local maxima. Hence, following their recommendation, the centre part is maximised from a grid of
starting values and the largest log-likelihood is
chosen. 
Standard errors are obtained from the inverse of the observed information matrix.

\section{Simulation} \label{sec:simulation}

In this section, a simulation study is conducted to investigate the finite-sample performance of the maximum likelihood estimators for the proposed CCR-TWvM, CCR-BvM and CCR-KvM models. 
Two covariate scenarios are considered: one linear and one circular interval-valued covariate, and three linear and one circular interval-valued covariate. 

For each scenario, a dataset of size $n$ is generated as follows. At first, the centres and
ranges of the linear and circular interval-valued covariates are generated, giving the
covariate vectors $\mathbf{x}_i^C$ and $\mathbf{x}_i^R$. Thereafter, the range
$q_i$ of the response is generated from the truncated Weibull, reparameterised beta or
Kumaraswamy distribution, with its parameter determined by the range equation in~\eqref{model}.
Lastly, given $q_i$ and $\mathbf{x}_i^C$, the centre $c_i$ is generated from the von Mises
distribution in~\eqref{eq:vm}, with mean direction $\mu_i$ from the centre equation
in~\eqref{model} and concentration $\upsilon(q_i)$. The bounds of each interval then follow from the centre and range.

\begin{figure}[H]
    \centering
    \includegraphics[width=\linewidth]{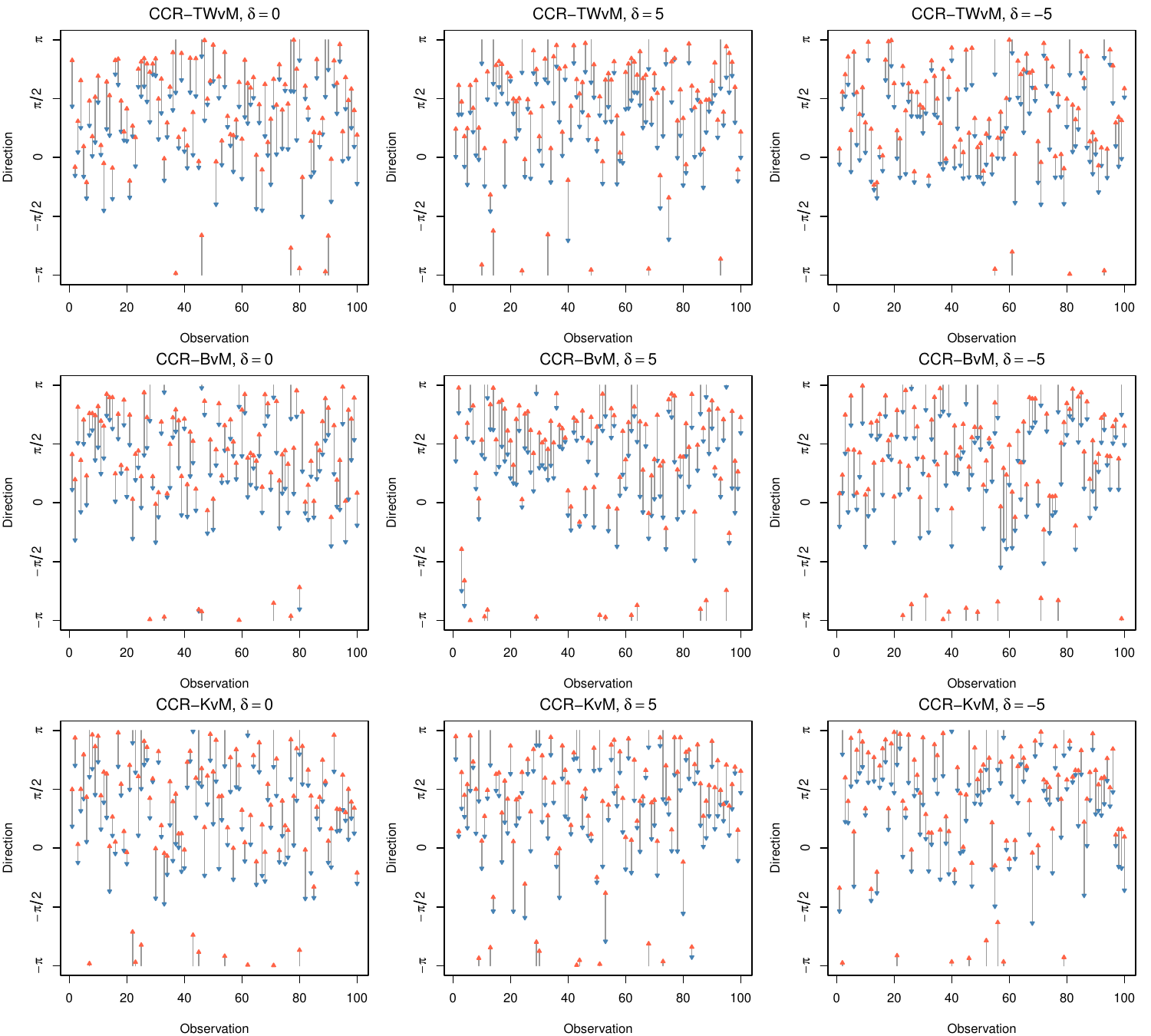}
    \caption{ Simulated circular interval-valued responses from the CCR-TWvM (top),
CCR-BvM (middle) and CCR-KvM (bottom) models with one linear and one circular covariate for $n=100$.}
    \label{fig:sims}
\end{figure}

For each configuration, sample sizes of \(n=100,500,1000\) are considered, with dependence parameter values \(\delta\in\{-5,0,5\}\). Each scenario is replicated 1000 times. Figure~\ref{fig:sims} illustrates the simulated responses for each of the scenarios in Tables ~\ref{tabTW} - \ref{tabK4}.

To reduce the risk of convergence to a local maximum, the centre component is estimated using multiple initial values, and the solution with the largest attained log-likelihood is retained. The finite-sample performance of the estimators is assessed using standard deviation, mean squared error (MSE) and coverage probabilities of 95\% confidence intervals, which are given in
Tables ~\ref{tabTW} - \ref{tabK4}. 

\begin{landscape}
\begin{figure}[H]
\centering
\includegraphics[width=\linewidth,height=0.28\textheight,keepaspectratio]{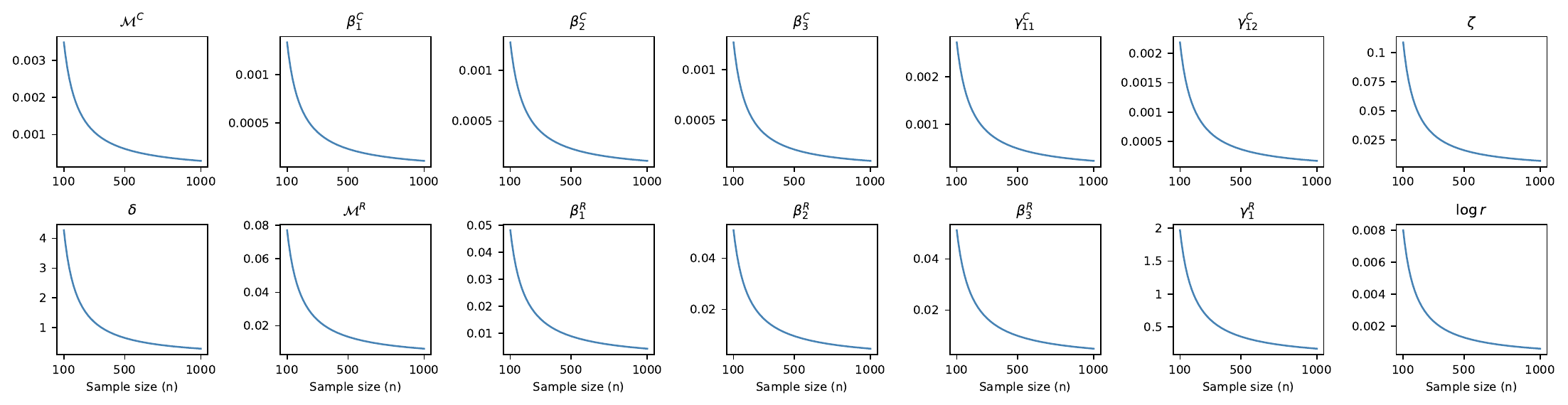}\\
\includegraphics[width=\linewidth,height=0.28\textheight,keepaspectratio]{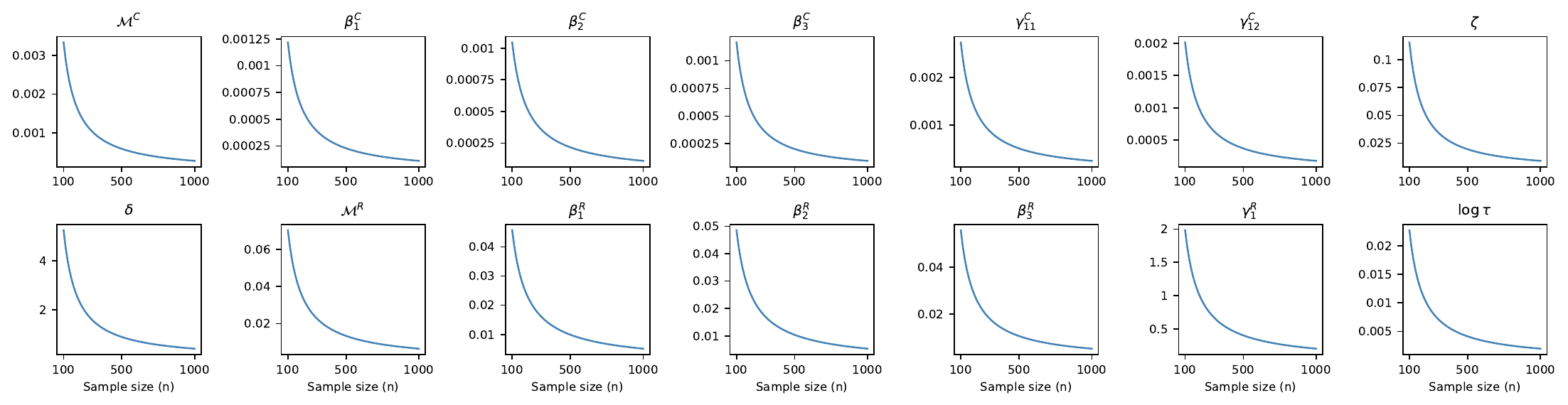}\\
\includegraphics[width=\linewidth,height=0.28\textheight,keepaspectratio]{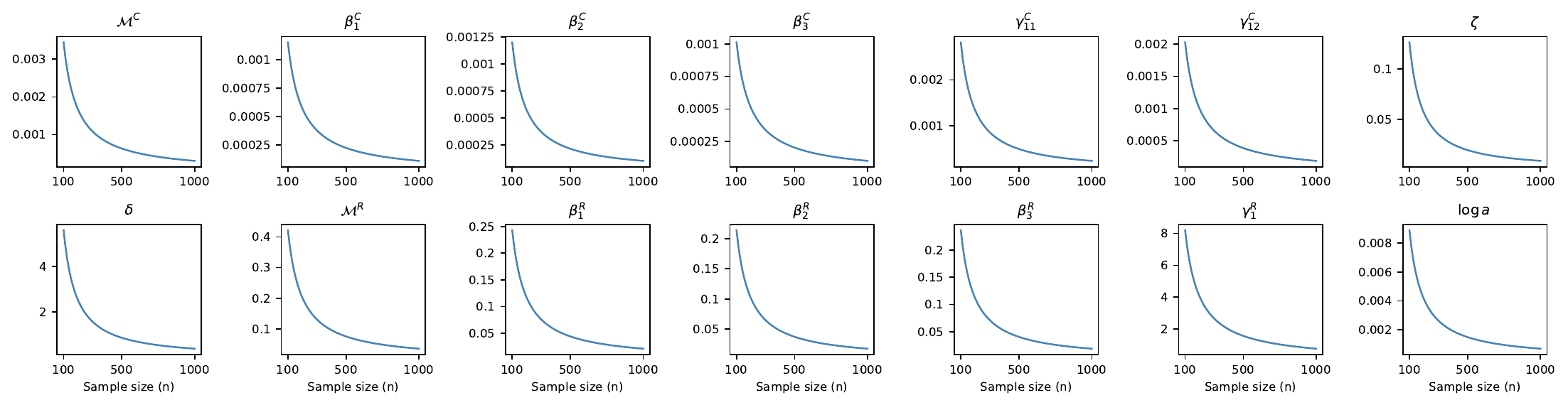}
\caption{\footnotesize MSE of the parameter estimators of the CCR-TWvM (top two rows), CCR-BvM (middle two rows)
and CCR-KvM (bottom two rows) models for three linear and one circular covariate and $\delta=5$.}
\label{fig:simintervals}
\end{figure}
\end{landscape}

For visualisation of the MSE over different sample
sizes, Figure~\ref{fig:simintervals} shows the MSE of the parameter estimators of the CCR-TWvM, CCR-BvM and
CCR-KvM models for different sample sizes and for the setting with three linear and one circular covariate where $\delta=5$.

The results in Tables~\ref{tabTW}--\ref{tabK4} show that the
maximum likelihood estimators perform well for all three range distributions and for different covariate scenarios. The empirical coverage probabilities are close to the nominal 95\% level, although for smaller sample sizes it is slightly smaller. The results in Figure~\ref{fig:simintervals} also demonstrate that the MSE of all the parameter estimators decreases as the sample size increases, for all three models. In conclusion, the results confirm the accuracy of the proposed maximum likelihood procedure for estimating the CCR models.

\begin{table}[H]
\centering
\caption{\footnotesize Parameter estimates, SDs, MSEs and CPs for CCR-TWvM model with one linear and one circular covariate.}
\label{tabTW}
\fbox{
\resizebox{0.9\linewidth}{!}{%
\begin{tabular}{llc cccc cccc cccc}
& & & \multicolumn{4}{c}{$\delta=0$} & \multicolumn{4}{c}{$\delta=5$} & \multicolumn{4}{c}{$\delta=-5$} \\
\cmidrule(lr){4-7} \cmidrule(lr){8-11} \cmidrule(lr){12-15}
Sample size & Parameter & True & Estimate & SD & MSE & CP & Estimate & SD & MSE & CP & Estimate & SD & MSE & CP \\
\midrule
\multirow{10}{*}{100} & $\mathcal{M}^C$ & 1.0000 & 0.9998 & 0.0401 & 0.0016 & 0.9440 & 1.0032 & 0.0600 & 0.0036 & 0.9140 & 1.0013 & 0.0280 & 0.0008 & 0.9380 \\
 & $\beta_1^C$ & 0.6000 & 0.6021 & 0.0304 & 0.0009 & 0.9280 & 0.6014 & 0.0423 & 0.0018 & 0.9360 & 0.6007 & 0.0205 & 0.0004 & 0.9360 \\
 & $\gamma_{11}^C$ & 0.4000 & 0.3996 & 0.0370 & 0.0014 & 0.9480 & 0.3962 & 0.0504 & 0.0026 & 0.9420 & 0.3984 & 0.0252 & 0.0006 & 0.9420 \\
 & $\gamma_{12}^C$ & -0.3000 & -0.3009 & 0.0334 & 0.0011 & 0.9280 & -0.2959 & 0.0420 & 0.0018 & 0.9660 & -0.2994 & 0.0239 & 0.0006 & 0.9200 \\
 & $\zeta$ & 2.3026 & 2.3596 & 0.3046 & 0.0959 & 0.9320 & 2.3691 & 0.3003 & 0.0944 & 0.9420 & 2.3241 & 0.3199 & 0.1026 & 0.9380 \\
 & $\delta$ & -- & -0.0595 & 2.0951 & 4.3843 & 0.9220 & 5.0820 & 1.9173 & 3.6754 & 0.9420 & -5.3033 & 2.1911 & 4.8832 & 0.9240 \\
 & $\mathcal{M}^R$ & -2.1400 & -2.1478 & 0.1980 & 0.0392 & 0.9200 & -2.1374 & 0.1918 & 0.0367 & 0.9340 & -2.1317 & 0.1793 & 0.0322 & 0.9320 \\
 & $\beta_1^R$ & 0.8000 & 0.7935 & 0.2244 & 0.0503 & 0.9460 & 0.7909 & 0.2215 & 0.0491 & 0.9460 & 0.8027 & 0.2274 & 0.0516 & 0.9220 \\
 & $\gamma_1^R$ & -3.0000 & -2.9462 & 1.4312 & 2.0470 & 0.9280 & -2.9764 & 1.4230 & 2.0215 & 0.9160 & -3.1567 & 1.3488 & 1.8402 & 0.9300 \\
 & $\log r$ & 0.6931 & 0.7152 & 0.0809 & 0.0070 & 0.9360 & 0.7237 & 0.0805 & 0.0074 & 0.9380 & 0.7180 & 0.0722 & 0.0058 & 0.9500 \\
\midrule
\multirow{10}{*}{500} & $\mathcal{M}^C$ & 1.0000 & 0.9998 & 0.0183 & 0.0003 & 0.9540 & 1.0004 & 0.0256 & 0.0007 & 0.9400 & 0.9997 & 0.0121 & 0.0001 & 0.9500 \\
 & $\beta_1^C$ & 0.6000 & 0.6008 & 0.0124 & 0.0002 & 0.9580 & 0.6004 & 0.0187 & 0.0004 & 0.9320 & 0.5996 & 0.0085 & 0.0001 & 0.9460 \\
 & $\gamma_{11}^C$ & 0.4000 & 0.4002 & 0.0145 & 0.0002 & 0.9720 & 0.3998 & 0.0229 & 0.0005 & 0.9340 & 0.4001 & 0.0112 & 0.0001 & 0.9400 \\
 & $\gamma_{12}^C$ & -0.3000 & -0.3011 & 0.0145 & 0.0002 & 0.9420 & -0.2998 & 0.0199 & 0.0004 & 0.9420 & -0.3001 & 0.0097 & 0.0001 & 0.9420 \\
 & $\zeta$ & 2.3026 & 2.3047 & 0.1211 & 0.0146 & 0.9620 & 2.3075 & 0.1235 & 0.0152 & 0.9360 & 2.2959 & 0.1240 & 0.0154 & 0.9660 \\
 & $\delta$ & 0 & -0.0640 & 0.7907 & 0.6281 & 0.9500 & 4.9733 & 0.7961 & 0.6332 & 0.9480 & -5.1144 & 0.8219 & 0.6872 & 0.9340 \\
 & $\mathcal{M}^R$ & -2.1400 & -2.1407 & 0.0843 & 0.0071 & 0.9360 & -2.1458 & 0.0812 & 0.0066 & 0.9500 & -2.1393 & 0.0774 & 0.0060 & 0.9620 \\
 & $\beta_1^R$ & 0.8000 & 0.8027 & 0.0992 & 0.0098 & 0.9440 & 0.8035 & 0.1009 & 0.0102 & 0.9400 & 0.8004 & 0.0947 & 0.0090 & 0.9480 \\
 & $\gamma_1^R$ & -3.0000 & -3.0041 & 0.5939 & 0.3521 & 0.9600 & -2.9682 & 0.6051 & 0.3664 & 0.9400 & -3.0136 & 0.5893 & 0.3468 & 0.9500 \\
 & $\log r$ & 0.6931 & 0.6987 & 0.0344 & 0.0012 & 0.9440 & 0.6971 & 0.0338 & 0.0012 & 0.9600 & 0.6999 & 0.0355 & 0.0013 & 0.9520 \\
\midrule
\multirow{10}{*}{1000} & $\mathcal{M}^C$ & 1.0000 & 1.0008 & 0.0130 & 0.0002 & 0.9420 & 1.0002 & 0.0168 & 0.0003 & 0.9640 & 1.0005 & 0.0085 & 0.0001 & 0.9480 \\
 & $\beta_1^C$ & 0.6000 & 0.6001 & 0.0092 & 0.0001 & 0.9460 & 0.5999 & 0.0125 & 0.0002 & 0.9580 & 0.5993 & 0.0059 & 0.0000 & 0.9620 \\
 & $\gamma_{11}^C$ & 0.4000 & 0.3992 & 0.0106 & 0.0001 & 0.9620 & 0.4001 & 0.0155 & 0.0002 & 0.9600 & 0.3997 & 0.0078 & 0.0001 & 0.9420 \\
 & $\gamma_{12}^C$ & -0.3000 & -0.3003 & 0.0102 & 0.0001 & 0.9400 & -0.2993 & 0.0147 & 0.0002 & 0.9400 & -0.2999 & 0.0068 & 0.0000 & 0.9540 \\
 & $\zeta$ & 2.3026 & 2.3054 & 0.0910 & 0.0083 & 0.9540 & 2.3114 & 0.0877 & 0.0078 & 0.9460 & 2.2994 & 0.0865 & 0.0075 & 0.9560 \\
 & $\delta$ & 0 & -0.0238 & 0.5942 & 0.3529 & 0.9400 & 5.0217 & 0.5447 & 0.2966 & 0.9520 & -5.0626 & 0.5509 & 0.3068 & 0.9560 \\
 & $\mathcal{M}^R$ & -2.1400 & -2.1379 & 0.0559 & 0.0031 & 0.9560 & -2.1401 & 0.0559 & 0.0031 & 0.9420 & -2.1382 & 0.0578 & 0.0033 & 0.9440 \\
 & $\beta_1^R$ & 0.8000 & 0.7966 & 0.0694 & 0.0048 & 0.9400 & 0.7967 & 0.0671 & 0.0045 & 0.9620 & 0.7991 & 0.0707 & 0.0050 & 0.9420 \\
 & $\gamma_1^R$ & -3.0000 & -2.9940 & 0.4167 & 0.1733 & 0.9580 & -2.9964 & 0.4355 & 0.1893 & 0.9520 & -2.9939 & 0.4260 & 0.1811 & 0.9420 \\
 & $\log r$ & 0.6931 & 0.6958 & 0.0246 & 0.0006 & 0.9580 & 0.6954 & 0.0249 & 0.0006 & 0.9400 & 0.6970 & 0.0243 & 0.0006 & 0.9420 \\
\end{tabular}}}
\end{table}

\begin{table}[H]
\centering
\caption{\footnotesize Parameter estimates, SDs, MSEs and CPs for CCR-BvM model with one linear and one circular covariate.}
\label{tabB}
\fbox{
\resizebox{0.9\linewidth}{!}{%
\begin{tabular}{llc cccc cccc cccc}
& & & \multicolumn{4}{c}{$\delta=0$} & \multicolumn{4}{c}{$\delta=5$} & \multicolumn{4}{c}{$\delta=-5$} \\
\cmidrule(lr){4-7} \cmidrule(lr){8-11} \cmidrule(lr){12-15}
Sample size & Parameter & True & Estimate & SD & MSE & CP & Estimate & SD & MSE & CP & Estimate & SD & MSE & CP \\
\midrule
\multirow{10}{*}{100} & $\mathcal{M}^C$ & 1.0000 & 1.0003 & 0.0405 & 0.0016 & 0.9560 & 0.9970 & 0.0584 & 0.0034 & 0.9400 & 0.9998 & 0.0280 & 0.0008 & 0.9460 \\
 & $\beta_1^C$ & 0.6000 & 0.6014 & 0.0286 & 0.0008 & 0.9480 & 0.6025 & 0.0412 & 0.0017 & 0.9380 & 0.6032 & 0.0225 & 0.0005 & 0.9200 \\
 & $\gamma_{11}^C$ & 0.4000 & 0.3990 & 0.0372 & 0.0014 & 0.9440 & 0.3998 & 0.0508 & 0.0026 & 0.9500 & 0.4005 & 0.0262 & 0.0007 & 0.9300 \\
 & $\gamma_{12}^C$ & -0.3000 & -0.3004 & 0.0333 & 0.0011 & 0.9320 & -0.3033 & 0.0447 & 0.0020 & 0.9320 & -0.3017 & 0.0229 & 0.0005 & 0.9280 \\
 & $\zeta$ & 2.3026 & 2.3191 & 0.3339 & 0.1115 & 0.9420 & 2.3815 & 0.3187 & 0.1076 & 0.9400 & 2.2880 & 0.3742 & 0.1399 & 0.9340 \\
 & $\delta$ & 0 & -0.2864 & 2.4581 & 6.1122 & 0.9260 & 5.2153 & 2.2054 & 4.9006 & 0.9520 & -5.5659 & 2.6520 & 7.3392 & 0.9160 \\
 & $\mathcal{M}^R$ & -2.1500 & -2.1631 & 0.1926 & 0.0372 & 0.9420 & -2.1481 & 0.1916 & 0.0367 & 0.9560 & -2.1642 & 0.1976 & 0.0392 & 0.9480 \\
 & $\beta_1^R$ & 0.8000 & 0.8142 & 0.2332 & 0.0545 & 0.9400 & 0.7961 & 0.2338 & 0.0546 & 0.9300 & 0.8096 & 0.2397 & 0.0575 & 0.9380 \\
 & $\gamma_1^R$ & -3.0000 & -2.9688 & 1.4704 & 2.1587 & 0.9360 & -3.0145 & 1.3824 & 1.9074 & 0.9440 & -2.9453 & 1.3138 & 1.7256 & 0.9620 \\
 & $\log\tau$ & 3.4012 & 3.4394 & 0.1421 & 0.0216 & 0.9320 & 3.4537 & 0.1383 & 0.0219 & 0.9480 & 3.4455 & 0.1457 & 0.0232 & 0.9280 \\
\midrule
\multirow{10}{*}{500} & $\mathcal{M}^C$ & 1.0000 & 0.9998 & 0.0178 & 0.0003 & 0.9540 & 0.9986 & 0.0245 & 0.0006 & 0.9560 & 1.0003 & 0.0120 & 0.0001 & 0.9480 \\
 & $\beta_1^C$ & 0.6000 & 0.6007 & 0.0133 & 0.0002 & 0.9300 & 0.6007 & 0.0184 & 0.0003 & 0.9380 & 0.5998 & 0.0087 & 0.0001 & 0.9520 \\
 & $\gamma_{11}^C$ & 0.4000 & 0.4001 & 0.0156 & 0.0002 & 0.9540 & 0.4018 & 0.0223 & 0.0005 & 0.9420 & 0.4001 & 0.0107 & 0.0001 & 0.9600 \\
 & $\gamma_{12}^C$ & -0.3000 & -0.3000 & 0.0140 & 0.0002 & 0.9460 & -0.3014 & 0.0202 & 0.0004 & 0.9320 & -0.2999 & 0.0101 & 0.0001 & 0.9420 \\
 & $\zeta$ & 2.3026 & 2.3123 & 0.1400 & 0.0197 & 0.9440 & 2.3218 & 0.1400 & 0.0199 & 0.9300 & 2.3043 & 0.1435 & 0.0206 & 0.9440 \\
 & $\delta$ & 0 & -0.0097 & 0.9694 & 0.9380 & 0.9580 & 5.0695 & 0.9237 & 0.8563 & 0.9460 & -5.0745 & 1.0011 & 1.0058 & 0.9300 \\
 & $\mathcal{M}^R$ & -2.1500 & -2.1483 & 0.0842 & 0.0071 & 0.9420 & -2.1515 & 0.0835 & 0.0070 & 0.9620 & -2.1451 & 0.0897 & 0.0081 & 0.9340 \\
 & $\beta_1^R$ & 0.8000 & 0.8045 & 0.0998 & 0.0100 & 0.9480 & 0.8041 & 0.1014 & 0.0103 & 0.9460 & 0.7977 & 0.1025 & 0.0105 & 0.9320 \\
 & $\gamma_1^R$ & -3.0000 & -3.0465 & 0.6307 & 0.3992 & 0.9400 & -3.0318 & 0.5998 & 0.3601 & 0.9560 & -3.0451 & 0.6098 & 0.3732 & 0.9480 \\
 & $\log\tau$ & 3.4012 & 3.4107 & 0.0632 & 0.0041 & 0.9480 & 3.4078 & 0.0615 & 0.0038 & 0.9580 & 3.4062 & 0.0626 & 0.0039 & 0.9480 \\
\midrule
\multirow{10}{*}{1000} & $\mathcal{M}^C$ & 1.0000 & 1.0000 & 0.0128 & 0.0002 & 0.9360 & 1.0006 & 0.0170 & 0.0003 & 0.9580 & 0.9999 & 0.0092 & 0.0001 & 0.9460 \\
 & $\beta_1^C$ & 0.6000 & 0.5997 & 0.0090 & 0.0001 & 0.9620 & 0.5995 & 0.0123 & 0.0002 & 0.9480 & 0.6001 & 0.0066 & 0.0000 & 0.9400 \\
 & $\gamma_{11}^C$ & 0.4000 & 0.3995 & 0.0115 & 0.0001 & 0.9460 & 0.3998 & 0.0156 & 0.0002 & 0.9360 & 0.4002 & 0.0084 & 0.0001 & 0.9420 \\
 & $\gamma_{12}^C$ & -0.3000 & -0.3002 & 0.0101 & 0.0001 & 0.9500 & -0.2996 & 0.0137 & 0.0002 & 0.9460 & -0.3004 & 0.0068 & 0.0000 & 0.9500 \\
 & $\zeta$ & 2.3026 & 2.3007 & 0.1001 & 0.0100 & 0.9320 & 2.3110 & 0.0976 & 0.0096 & 0.9400 & 2.3041 & 0.0979 & 0.0096 & 0.9540 \\
 & $\delta$ & 0 & -0.0609 & 0.6883 & 0.4765 & 0.9440 & 5.0077 & 0.6554 & 0.4287 & 0.9580 & -5.0372 & 0.6787 & 0.4611 & 0.9540 \\
 & $\mathcal{M}^R$ & -2.1500 & -2.1524 & 0.0594 & 0.0035 & 0.9460 & -2.1500 & 0.0629 & 0.0039 & 0.9400 & -2.1538 & 0.0599 & 0.0036 & 0.9480 \\
 & $\beta_1^R$ & 0.8000 & 0.8050 & 0.0719 & 0.0052 & 0.9420 & 0.7967 & 0.0757 & 0.0057 & 0.9280 & 0.8016 & 0.0709 & 0.0050 & 0.9500 \\
 & $\gamma_1^R$ & -3.0000 & -3.0135 & 0.4237 & 0.1793 & 0.9460 & -2.9961 & 0.4231 & 0.1786 & 0.9420 & -2.9745 & 0.4445 & 0.1979 & 0.9440 \\
 & $\log\tau$ & 3.4012 & 3.4077 & 0.0458 & 0.0021 & 0.9500 & 3.4057 & 0.0461 & 0.0021 & 0.9420 & 3.4032 & 0.0477 & 0.0023 & 0.9240 \\
\end{tabular}}}
\end{table}

\begin{table}[H]
\centering
\caption{\footnotesize Parameter estimates, SDs, MSEs and CPs for CCR-KvM model with one linear and one circular covariate.}
\label{tabK}
\fbox{
\resizebox{0.9\linewidth}{!}{%
\begin{tabular}{llc cccc cccc cccc}
& & & \multicolumn{4}{c}{$\delta=0$} & \multicolumn{4}{c}{$\delta=5$} & \multicolumn{4}{c}{$\delta=-5$} \\
\cmidrule(lr){4-7} \cmidrule(lr){8-11} \cmidrule(lr){12-15}
Sample size & Parameter & True & Estimate & SD & MSE & CP & Estimate & SD & MSE & CP & Estimate & SD & MSE & CP \\
\midrule
\multirow{10}{*}{100} & $\mathcal{M}^C$ & 1.0000 & 1.0000 & 0.0435 & 0.0019 & 0.9260 & 0.9997 & 0.0572 & 0.0033 & 0.9460 & 0.9993 & 0.0285 & 0.0008 & 0.9440 \\
 & $\beta_1^C$ & 0.6000 & 0.5989 & 0.0284 & 0.0008 & 0.9480 & 0.6009 & 0.0413 & 0.0017 & 0.9380 & 0.5987 & 0.0204 & 0.0004 & 0.9340 \\
 & $\gamma_{11}^C$ & 0.4000 & 0.4013 & 0.0356 & 0.0013 & 0.9460 & 0.4017 & 0.0517 & 0.0027 & 0.9420 & 0.4015 & 0.0265 & 0.0007 & 0.9320 \\
 & $\gamma_{12}^C$ & -0.3000 & -0.2978 & 0.0318 & 0.0010 & 0.9380 & -0.3025 & 0.0469 & 0.0022 & 0.9220 & -0.3014 & 0.0226 & 0.0005 & 0.9380 \\
 & $\zeta$ & 2.3026 & 2.3746 & 0.3136 & 0.1033 & 0.9480 & 2.3751 & 0.3008 & 0.0955 & 0.9500 & 2.3060 & 0.3275 & 0.1071 & 0.9400 \\
 & $\delta$ & 0 & 0.0326 & 2.2584 & 5.0913 & 0.9460 & 5.1535 & 2.0921 & 4.3918 & 0.9620 & -5.4252 & 2.3167 & 5.5370 & 0.9380 \\
 & $\mathcal{M}^R$ & 4.0800 & 4.2126 & 0.4914 & 0.2586 & 0.9480 & 4.1935 & 0.4920 & 0.2545 & 0.9520 & 4.2237 & 0.4771 & 0.2478 & 0.9460 \\
 & $\beta_1^R$ & -0.8000 & -0.8163 & 0.4586 & 0.2101 & 0.9320 & -0.8543 & 0.4501 & 0.2051 & 0.9400 & -0.8277 & 0.4622 & 0.2140 & 0.9300 \\
 & $\gamma_1^R$ & 3.0000 & 2.9260 & 2.8673 & 8.2104 & 0.9380 & 3.3038 & 2.8473 & 8.1832 & 0.9500 & 2.8946 & 2.9201 & 8.5209 & 0.9380 \\
 & $\log a$ & 0.6931 & 0.7200 & 0.0824 & 0.0075 & 0.9400 & 0.7131 & 0.0823 & 0.0072 & 0.9440 & 0.7182 & 0.0857 & 0.0080 & 0.9280 \\
\midrule
\multirow{10}{*}{500} & $\mathcal{M}^C$ & 1.0000 & 0.9998 & 0.0187 & 0.0003 & 0.9380 & 1.0002 & 0.0257 & 0.0007 & 0.9360 & 1.0005 & 0.0121 & 0.0001 & 0.9520 \\
 & $\beta_1^C$ & 0.6000 & 0.6000 & 0.0126 & 0.0002 & 0.9560 & 0.6005 & 0.0186 & 0.0003 & 0.9360 & 0.6005 & 0.0089 & 0.0001 & 0.9460 \\
 & $\gamma_{11}^C$ & 0.4000 & 0.3997 & 0.0167 & 0.0003 & 0.9340 & 0.3997 & 0.0223 & 0.0005 & 0.9500 & 0.4003 & 0.0112 & 0.0001 & 0.9420 \\
 & $\gamma_{12}^C$ & -0.3000 & -0.2995 & 0.0134 & 0.0002 & 0.9600 & -0.3020 & 0.0202 & 0.0004 & 0.9400 & -0.3006 & 0.0095 & 0.0001 & 0.9560 \\
 & $\zeta$ & 2.3026 & 2.3143 & 0.1366 & 0.0187 & 0.9400 & 2.3211 & 0.1305 & 0.0173 & 0.9360 & 2.3107 & 0.1398 & 0.0196 & 0.9440 \\
 & $\delta$ & 0 & -0.0234 & 0.9212 & 0.8474 & 0.9460 & 5.0842 & 0.9023 & 0.8196 & 0.9320 & -5.0102 & 0.9635 & 0.9266 & 0.9520 \\
 & $\mathcal{M}^R$ & 4.0800 & 4.1105 & 0.2070 & 0.0437 & 0.9480 & 4.0933 & 0.2195 & 0.0482 & 0.9480 & 4.1074 & 0.2131 & 0.0461 & 0.9680 \\
 & $\beta_1^R$ & -0.8000 & -0.8212 & 0.1894 & 0.0362 & 0.9600 & -0.8012 & 0.1985 & 0.0393 & 0.9460 & -0.8030 & 0.1999 & 0.0399 & 0.9420 \\
 & $\gamma_1^R$ & 3.0000 & 3.0190 & 1.2344 & 1.5209 & 0.9420 & 3.0464 & 1.1958 & 1.4292 & 0.9560 & 3.0266 & 1.1988 & 1.4349 & 0.9540 \\
 & $\log a$ & 0.6931 & 0.6969 & 0.0367 & 0.0014 & 0.9460 & 0.6968 & 0.0381 & 0.0015 & 0.9400 & 0.6992 & 0.0364 & 0.0014 & 0.9480 \\
\midrule
\multirow{10}{*}{1000} & $\mathcal{M}^C$ & 1.0000 & 0.9994 & 0.0123 & 0.0002 & 0.9600 & 1.0009 & 0.0172 & 0.0003 & 0.9560 & 1.0002 & 0.0090 & 0.0001 & 0.9380 \\
 & $\beta_1^C$ & 0.6000 & 0.6002 & 0.0091 & 0.0001 & 0.9440 & 0.6008 & 0.0117 & 0.0001 & 0.9520 & 0.5999 & 0.0064 & 0.0000 & 0.9400 \\
 & $\gamma_{11}^C$ & 0.4000 & 0.4007 & 0.0112 & 0.0001 & 0.9500 & 0.4011 & 0.0147 & 0.0002 & 0.9620 & 0.4003 & 0.0080 & 0.0001 & 0.9360 \\
 & $\gamma_{12}^C$ & -0.3000 & -0.2999 & 0.0107 & 0.0001 & 0.9400 & -0.3005 & 0.0139 & 0.0002 & 0.9560 & -0.3003 & 0.0073 & 0.0001 & 0.9300 \\
 & $\zeta$ & 2.3026 & 2.3077 & 0.0968 & 0.0094 & 0.9460 & 2.3104 & 0.0908 & 0.0083 & 0.9460 & 2.3014 & 0.0975 & 0.0095 & 0.9440 \\
 & $\delta$ & 0 & 0.0187 & 0.6575 & 0.4318 & 0.9500 & 5.0073 & 0.6098 & 0.3711 & 0.9480 & -5.0780 & 0.6864 & 0.4763 & 0.9340 \\
 & $\mathcal{M}^R$ & 4.0800 & 4.0947 & 0.1536 & 0.0238 & 0.9360 & 4.0878 & 0.1501 & 0.0225 & 0.9500 & 4.0943 & 0.1472 & 0.0218 & 0.9560 \\
 & $\beta_1^R$ & -0.8000 & -0.8051 & 0.1418 & 0.0201 & 0.9500 & -0.8007 & 0.1361 & 0.0185 & 0.9560 & -0.8000 & 0.1376 & 0.0189 & 0.9520 \\
 & $\gamma_1^R$ & 3.0000 & 3.0452 & 0.8639 & 0.7469 & 0.9340 & 3.0578 & 0.8508 & 0.7258 & 0.9380 & 3.0331 & 0.8557 & 0.7318 & 0.9540 \\
 & $\log a$ & 0.6931 & 0.6966 & 0.0254 & 0.0007 & 0.9580 & 0.6956 & 0.0261 & 0.0007 & 0.9420 & 0.6973 & 0.0254 & 0.0007 & 0.9520 \\
\end{tabular}}}
\end{table}

\begin{table}[H]
\centering
\caption{\footnotesize Parameter estimates, SDs, MSEs and CPs for CCR-TWvM model with three linear and one circular covariate.}
\label{tabTW4}
\fbox{
\resizebox{0.9\linewidth}{!}{%
\begin{tabular}{llc cccc cccc cccc}
& & & \multicolumn{4}{c}{$\delta=0$} & \multicolumn{4}{c}{$\delta=5$} & \multicolumn{4}{c}{$\delta=-5$} \\
\cmidrule(lr){4-7} \cmidrule(lr){8-11} \cmidrule(lr){12-15}
Sample size & Parameter & True & Estimate & SD & MSE & CP & Estimate & SD & MSE & CP & Estimate & SD & MSE & CP \\
\midrule
\multirow{14}{*}{100} & $\mathcal{M}^C$ & 1.0000 & 0.9999 & 0.0402 & 0.0016 & 0.9500 & 1.0008 & 0.0602 & 0.0036 & 0.9160 & 0.9976 & 0.0292 & 0.0009 & 0.9240 \\
 & $\beta_1^C$ & 0.4000 & 0.4015 & 0.0264 & 0.0007 & 0.9380 & 0.4008 & 0.0362 & 0.0013 & 0.9440 & 0.3986 & 0.0174 & 0.0003 & 0.9360 \\
 & $\beta_2^C$ & -0.3500 & -0.3513 & 0.0248 & 0.0006 & 0.9440 & -0.3504 & 0.0362 & 0.0013 & 0.9020 & -0.3518 & 0.0172 & 0.0003 & 0.9180 \\
 & $\beta_3^C$ & 0.3000 & 0.3023 & 0.0224 & 0.0005 & 0.9480 & 0.3005 & 0.0356 & 0.0013 & 0.9100 & 0.2996 & 0.0161 & 0.0003 & 0.9380 \\
 & $\gamma_{11}^C$ & 0.4000 & 0.4009 & 0.0353 & 0.0012 & 0.9580 & 0.4016 & 0.0524 & 0.0027 & 0.9300 & 0.4001 & 0.0258 & 0.0007 & 0.9300 \\
 & $\gamma_{12}^C$ & -0.3000 & -0.3011 & 0.0337 & 0.0011 & 0.9140 & -0.2975 & 0.0464 & 0.0022 & 0.9200 & -0.2982 & 0.0237 & 0.0006 & 0.9320 \\
 & $\zeta$ & 2.3026 & 2.3811 & 0.3101 & 0.1021 & 0.9320 & 2.4311 & 0.3049 & 0.1093 & 0.9200 & 2.3357 & 0.3543 & 0.1264 & 0.9120 \\
 & $\delta$ & 0 & -0.0515 & 2.1094 & 4.4434 & 0.9460 & 5.3410 & 2.0455 & 4.2919 & 0.9300 & -5.3899 & 2.5016 & 6.3976 & 0.8860 \\
 & $\mathcal{M}^R$ & -1.9700 & -1.9724 & 0.2662 & 0.0707 & 0.9560 & -1.9841 & 0.2752 & 0.0758 & 0.9340 & -1.9805 & 0.2540 & 0.0645 & 0.9400 \\
 & $\beta_1^R$ & 0.5000 & 0.4942 & 0.2236 & 0.0499 & 0.9440 & 0.4880 & 0.2193 & 0.0481 & 0.9480 & 0.5065 & 0.2219 & 0.0492 & 0.9340 \\
 & $\beta_2^R$ & -0.4000 & -0.4031 & 0.2318 & 0.0536 & 0.9420 & -0.3875 & 0.2239 & 0.0502 & 0.9380 & -0.4161 & 0.2192 & 0.0482 & 0.9560 \\
 & $\beta_3^R$ & 0.4000 & 0.4076 & 0.2241 & 0.0502 & 0.9500 & 0.4132 & 0.2269 & 0.0515 & 0.9260 & 0.4193 & 0.2213 & 0.0493 & 0.9540 \\
 & $\gamma_1^R$ & -3.0000 & -3.0205 & 1.3629 & 1.8542 & 0.9500 & -3.0078 & 1.3906 & 1.9301 & 0.9500 & -3.0227 & 1.3653 & 1.8609 & 0.9380 \\
 & $\log r$ & 0.6931 & 0.7214 & 0.0780 & 0.0069 & 0.9360 & 0.7276 & 0.0822 & 0.0079 & 0.9180 & 0.7329 & 0.0775 & 0.0076 & 0.9320 \\
\midrule
\multirow{14}{*}{500} & $\mathcal{M}^C$ & 1.0000 & 0.9994 & 0.0181 & 0.0003 & 0.9520 & 1.0013 & 0.0232 & 0.0005 & 0.9660 & 0.9989 & 0.0123 & 0.0002 & 0.9480 \\
 & $\beta_1^C$ & 0.4000 & 0.4002 & 0.0108 & 0.0001 & 0.9580 & 0.4010 & 0.0158 & 0.0002 & 0.9440 & 0.3996 & 0.0073 & 0.0001 & 0.9500 \\
 & $\beta_2^C$ & -0.3500 & -0.3494 & 0.0108 & 0.0001 & 0.9460 & -0.3514 & 0.0145 & 0.0002 & 0.9460 & -0.3498 & 0.0075 & 0.0001 & 0.9400 \\
 & $\beta_3^C$ & 0.3000 & 0.3000 & 0.0105 & 0.0001 & 0.9320 & 0.2995 & 0.0145 & 0.0002 & 0.9460 & 0.3000 & 0.0074 & 0.0001 & 0.9300 \\
 & $\gamma_{11}^C$ & 0.4000 & 0.4010 & 0.0158 & 0.0003 & 0.9640 & 0.4001 & 0.0220 & 0.0005 & 0.9500 & 0.4009 & 0.0109 & 0.0001 & 0.9500 \\
 & $\gamma_{12}^C$ & -0.3000 & -0.3001 & 0.0141 & 0.0002 & 0.9520 & -0.3006 & 0.0197 & 0.0004 & 0.9560 & -0.3003 & 0.0095 & 0.0001 & 0.9560 \\
 & $\zeta$ & 2.3026 & 2.3165 & 0.1269 & 0.0163 & 0.9500 & 2.3095 & 0.1259 & 0.0159 & 0.9420 & 2.3113 & 0.1266 & 0.0161 & 0.9620 \\
 & $\delta$ & 0 & -0.0359 & 0.8189 & 0.6706 & 0.9560 & 4.9924 & 0.8037 & 0.6447 & 0.9400 & -5.0668 & 0.8356 & 0.7014 & 0.9660 \\
 & $\mathcal{M}^R$ & -1.9700 & -1.9737 & 0.1199 & 0.0144 & 0.9400 & -1.9634 & 0.1195 & 0.0143 & 0.9260 & -1.9705 & 0.1111 & 0.0123 & 0.9580 \\
 & $\beta_1^R$ & 0.5000 & 0.5028 & 0.0930 & 0.0086 & 0.9580 & 0.4944 & 0.0947 & 0.0090 & 0.9580 & 0.5044 & 0.0947 & 0.0090 & 0.9600 \\
 & $\beta_2^R$ & -0.4000 & -0.4023 & 0.0953 & 0.0091 & 0.9460 & -0.4001 & 0.1007 & 0.0101 & 0.9460 & -0.3973 & 0.0958 & 0.0092 & 0.9480 \\
 & $\beta_3^R$ & 0.4000 & 0.4032 & 0.0985 & 0.0097 & 0.9560 & 0.4000 & 0.0989 & 0.0098 & 0.9440 & 0.3988 & 0.1002 & 0.0100 & 0.9480 \\
 & $\gamma_1^R$ & -3.0000 & -2.9838 & 0.6122 & 0.3743 & 0.9400 & -3.0217 & 0.6142 & 0.3770 & 0.9540 & -3.0379 & 0.5970 & 0.3572 & 0.9620 \\
 & $\log r$ & 0.6931 & 0.7009 & 0.0347 & 0.0013 & 0.9360 & 0.6996 & 0.0357 & 0.0013 & 0.9360 & 0.7002 & 0.0349 & 0.0013 & 0.9460 \\
\midrule
\multirow{14}{*}{1000} & $\mathcal{M}^C$ & 1.0000 & 0.9998 & 0.0131 & 0.0002 & 0.9520 & 1.0008 & 0.0179 & 0.0003 & 0.9520 & 1.0004 & 0.0089 & 0.0001 & 0.9320 \\
 & $\beta_1^C$ & 0.4000 & 0.4003 & 0.0079 & 0.0001 & 0.9500 & 0.4003 & 0.0103 & 0.0001 & 0.9600 & 0.4001 & 0.0051 & 0.0000 & 0.9480 \\
 & $\beta_2^C$ & -0.3500 & -0.3500 & 0.0081 & 0.0001 & 0.9320 & -0.3509 & 0.0108 & 0.0001 & 0.9480 & -0.3502 & 0.0052 & 0.0000 & 0.9420 \\
 & $\beta_3^C$ & 0.3000 & 0.2992 & 0.0076 & 0.0001 & 0.9460 & 0.3008 & 0.0097 & 0.0001 & 0.9620 & 0.3000 & 0.0049 & 0.0000 & 0.9600 \\
 & $\gamma_{11}^C$ & 0.4000 & 0.4000 & 0.0120 & 0.0001 & 0.9340 & 0.3995 & 0.0156 & 0.0002 & 0.9500 & 0.4001 & 0.0079 & 0.0001 & 0.9360 \\
 & $\gamma_{12}^C$ & -0.3000 & -0.3014 & 0.0104 & 0.0001 & 0.9420 & -0.2998 & 0.0129 & 0.0002 & 0.9640 & -0.3000 & 0.0068 & 0.0000 & 0.9480 \\
 & $\zeta$ & 2.3026 & 2.3054 & 0.0924 & 0.0085 & 0.9340 & 2.3067 & 0.0852 & 0.0073 & 0.9560 & 2.3047 & 0.0918 & 0.0084 & 0.9400 \\
 & $\delta$ & 0 & -0.0160 & 0.6057 & 0.3664 & 0.9440 & 4.9821 & 0.5410 & 0.2924 & 0.9520 & -5.0546 & 0.5861 & 0.3458 & 0.9600 \\
 & $\mathcal{M}^R$ & -1.9700 & -1.9640 & 0.0829 & 0.0069 & 0.9560 & -1.9641 & 0.0781 & 0.0061 & 0.9620 & -1.9735 & 0.0858 & 0.0074 & 0.9440 \\
 & $\beta_1^R$ & 0.5000 & 0.4957 & 0.0680 & 0.0046 & 0.9440 & 0.4946 & 0.0655 & 0.0043 & 0.9580 & 0.4996 & 0.0711 & 0.0050 & 0.9500 \\
 & $\beta_2^R$ & -0.4000 & -0.4003 & 0.0693 & 0.0048 & 0.9460 & -0.4001 & 0.0671 & 0.0045 & 0.9540 & -0.3978 & 0.0683 & 0.0047 & 0.9480 \\
 & $\beta_3^R$ & 0.4000 & 0.3955 & 0.0689 & 0.0048 & 0.9400 & 0.3981 & 0.0705 & 0.0050 & 0.9420 & 0.4029 & 0.0674 & 0.0045 & 0.9400 \\
 & $\gamma_1^R$ & -3.0000 & -3.0123 & 0.3989 & 0.1590 & 0.9580 & -3.0189 & 0.4002 & 0.1602 & 0.9580 & -2.9923 & 0.4217 & 0.1776 & 0.9580 \\
 & $\log r$ & 0.6931 & 0.6971 & 0.0243 & 0.0006 & 0.9540 & 0.6957 & 0.0238 & 0.0006 & 0.9600 & 0.6964 & 0.0242 & 0.0006 & 0.9560 \\
\end{tabular}}}
\end{table}

\begin{table}[H]
\centering
\caption{\footnotesize Parameter estimates, SDs, MSEs and CPs for CCR-BvM model with three linear and one circular covariate.}
\label{tabB4}
\fbox{
\resizebox{0.9\linewidth}{!}{%
\begin{tabular}{llc cccc cccc cccc}
& & & \multicolumn{4}{c}{$\delta=0$} & \multicolumn{4}{c}{$\delta=5$} & \multicolumn{4}{c}{$\delta=-5$} \\
\cmidrule(lr){4-7} \cmidrule(lr){8-11} \cmidrule(lr){12-15}
Sample size & Parameter & True & Estimate & SD & MSE & CP & Estimate & SD & MSE & CP & Estimate & SD & MSE & CP \\
\midrule
\multirow{14}{*}{100} & $\mathcal{M}^C$ & 1.0000 & 0.9989 & 0.0414 & 0.0017 & 0.9340 & 1.0016 & 0.0581 & 0.0034 & 0.9420 & 1.0017 & 0.0288 & 0.0008 & 0.9460 \\
 & $\beta_1^C$ & 0.4000 & 0.3994 & 0.0250 & 0.0006 & 0.9440 & 0.3996 & 0.0352 & 0.0012 & 0.9460 & 0.4005 & 0.0172 & 0.0003 & 0.9520 \\
 & $\beta_2^C$ & -0.3500 & -0.3491 & 0.0259 & 0.0007 & 0.9260 & -0.3482 & 0.0324 & 0.0011 & 0.9480 & -0.3502 & 0.0184 & 0.0003 & 0.9240 \\
 & $\beta_3^C$ & 0.3000 & 0.2989 & 0.0252 & 0.0006 & 0.9320 & 0.3008 & 0.0341 & 0.0012 & 0.9400 & 0.3002 & 0.0169 & 0.0003 & 0.9360 \\
 & $\gamma_{11}^C$ & 0.4000 & 0.4003 & 0.0369 & 0.0014 & 0.9480 & 0.4009 & 0.0523 & 0.0027 & 0.9460 & 0.4000 & 0.0265 & 0.0007 & 0.9260 \\
 & $\gamma_{12}^C$ & -0.3000 & -0.3009 & 0.0323 & 0.0010 & 0.9500 & -0.2996 & 0.0451 & 0.0020 & 0.9340 & -0.2992 & 0.0234 & 0.0005 & 0.9180 \\
 & $\zeta$ & 2.3026 & 2.3805 & 0.3606 & 0.1358 & 0.9220 & 2.3883 & 0.3266 & 0.1138 & 0.9300 & 2.2927 & 0.3747 & 0.1402 & 0.9440 \\
 & $\delta$ & 0 & -0.1499 & 2.6261 & 6.9051 & 0.9260 & 5.1000 & 2.2672 & 5.1399 & 0.9460 & -5.6641 & 2.7324 & 7.8922 & 0.9160 \\
 & $\mathcal{M}^R$ & -1.9700 & -1.9727 & 0.2619 & 0.0684 & 0.9460 & -1.9677 & 0.2647 & 0.0699 & 0.9560 & -1.9888 & 0.2728 & 0.0746 & 0.9640 \\
 & $\beta_1^R$ & 0.5000 & 0.5180 & 0.2455 & 0.0605 & 0.9280 & 0.5062 & 0.2126 & 0.0451 & 0.9660 & 0.4936 & 0.2284 & 0.0521 & 0.9320 \\
 & $\beta_2^R$ & -0.4000 & -0.4059 & 0.2254 & 0.0507 & 0.9500 & -0.4044 & 0.2203 & 0.0484 & 0.9540 & -0.4000 & 0.2268 & 0.0513 & 0.9480 \\
 & $\beta_3^R$ & 0.4000 & 0.3992 & 0.2392 & 0.0571 & 0.9480 & 0.4011 & 0.2355 & 0.0554 & 0.9460 & 0.4216 & 0.2074 & 0.0434 & 0.9600 \\
 & $\gamma_1^R$ & -3.0000 & -3.0765 & 1.4321 & 2.0526 & 0.9420 & -3.0467 & 1.4264 & 2.0328 & 0.9420 & -2.9292 & 1.3712 & 1.8816 & 0.9540 \\
 & $\log\tau$ & 3.4012 & 3.4547 & 0.1321 & 0.0203 & 0.9480 & 3.4415 & 0.1446 & 0.0225 & 0.9280 & 3.4588 & 0.1407 & 0.0231 & 0.9260 \\
\midrule
\multirow{14}{*}{500} & $\mathcal{M}^C$ & 1.0000 & 0.9993 & 0.0174 & 0.0003 & 0.9520 & 1.0001 & 0.0240 & 0.0006 & 0.9580 & 1.0002 & 0.0116 & 0.0001 & 0.9540 \\
 & $\beta_1^C$ & 0.4000 & 0.3998 & 0.0110 & 0.0001 & 0.9440 & 0.4001 & 0.0148 & 0.0002 & 0.9600 & 0.4004 & 0.0076 & 0.0001 & 0.9500 \\
 & $\beta_2^C$ & -0.3500 & -0.3497 & 0.0107 & 0.0001 & 0.9480 & -0.3492 & 0.0146 & 0.0002 & 0.9460 & -0.3503 & 0.0070 & 0.0000 & 0.9680 \\
 & $\beta_3^C$ & 0.3000 & 0.2999 & 0.0099 & 0.0001 & 0.9600 & 0.3005 & 0.0143 & 0.0002 & 0.9500 & 0.3002 & 0.0071 & 0.0001 & 0.9520 \\
 & $\gamma_{11}^C$ & 0.4000 & 0.3992 & 0.0156 & 0.0002 & 0.9520 & 0.4009 & 0.0227 & 0.0005 & 0.9560 & 0.3999 & 0.0112 & 0.0001 & 0.9440 \\
 & $\gamma_{12}^C$ & -0.3000 & -0.2995 & 0.0136 & 0.0002 & 0.9620 & -0.2998 & 0.0190 & 0.0004 & 0.9580 & -0.3007 & 0.0097 & 0.0001 & 0.9440 \\
 & $\zeta$ & 2.3026 & 2.3118 & 0.1391 & 0.0194 & 0.9380 & 2.3206 & 0.1423 & 0.0205 & 0.9440 & 2.3106 & 0.1418 & 0.0201 & 0.9560 \\
 & $\delta$ & 0 & -0.0343 & 0.9706 & 0.9415 & 0.9420 & 5.0109 & 0.9738 & 0.9465 & 0.9440 & -5.1085 & 0.9965 & 1.0028 & 0.9480 \\
 & $\mathcal{M}^R$ & -1.9700 & -1.9701 & 0.1201 & 0.0144 & 0.9440 & -1.9672 & 0.1168 & 0.0136 & 0.9580 & -1.9753 & 0.1165 & 0.0136 & 0.9540 \\
 & $\beta_1^R$ & 0.5000 & 0.4973 & 0.0994 & 0.0099 & 0.9620 & 0.5010 & 0.1018 & 0.0103 & 0.9380 & 0.4946 & 0.1008 & 0.0102 & 0.9440 \\
 & $\beta_2^R$ & -0.4000 & -0.3997 & 0.0992 & 0.0098 & 0.9600 & -0.4044 & 0.1021 & 0.0104 & 0.9340 & -0.3988 & 0.0982 & 0.0096 & 0.9560 \\
 & $\beta_3^R$ & 0.4000 & 0.3994 & 0.1035 & 0.0107 & 0.9400 & 0.3951 & 0.1034 & 0.0107 & 0.9460 & 0.4034 & 0.1022 & 0.0104 & 0.9500 \\
 & $\gamma_1^R$ & -3.0000 & -3.0025 & 0.6438 & 0.4136 & 0.9420 & -2.9929 & 0.6085 & 0.3696 & 0.9480 & -2.9732 & 0.6081 & 0.3698 & 0.9460 \\
 & $\log\tau$ & 3.4012 & 3.4131 & 0.0630 & 0.0041 & 0.9540 & 3.4097 & 0.0647 & 0.0042 & 0.9420 & 3.4140 & 0.0640 & 0.0042 & 0.9320 \\
\midrule
\multirow{14}{*}{1000} & $\mathcal{M}^C$ & 1.0000 & 0.9997 & 0.0132 & 0.0002 & 0.9300 & 1.0005 & 0.0169 & 0.0003 & 0.9700 & 1.0002 & 0.0089 & 0.0001 & 0.9420 \\
 & $\beta_1^C$ & 0.4000 & 0.3999 & 0.0076 & 0.0001 & 0.9560 & 0.4003 & 0.0107 & 0.0001 & 0.9560 & 0.4000 & 0.0056 & 0.0000 & 0.9420 \\
 & $\beta_2^C$ & -0.3500 & -0.3502 & 0.0073 & 0.0001 & 0.9580 & -0.3497 & 0.0105 & 0.0001 & 0.9420 & -0.3499 & 0.0054 & 0.0000 & 0.9460 \\
 & $\beta_3^C$ & 0.3000 & 0.3003 & 0.0069 & 0.0000 & 0.9540 & 0.2993 & 0.0097 & 0.0001 & 0.9600 & 0.2999 & 0.0052 & 0.0000 & 0.9500 \\
 & $\gamma_{11}^C$ & 0.4000 & 0.4001 & 0.0113 & 0.0001 & 0.9460 & 0.4000 & 0.0156 & 0.0002 & 0.9520 & 0.3999 & 0.0080 & 0.0001 & 0.9440 \\
 & $\gamma_{12}^C$ & -0.3000 & -0.2995 & 0.0104 & 0.0001 & 0.9320 & -0.3007 & 0.0135 & 0.0002 & 0.9600 & -0.3001 & 0.0065 & 0.0000 & 0.9640 \\
 & $\zeta$ & 2.3026 & 2.3105 & 0.0997 & 0.0100 & 0.9480 & 2.3071 & 0.0933 & 0.0087 & 0.9480 & 2.3035 & 0.1018 & 0.0103 & 0.9460 \\
 & $\delta$ & 0 & -0.0102 & 0.6908 & 0.4764 & 0.9360 & 4.9971 & 0.6320 & 0.3986 & 0.9520 & -5.0758 & 0.7125 & 0.5124 & 0.9380 \\
 & $\mathcal{M}^R$ & -1.9700 & -1.9700 & 0.0848 & 0.0072 & 0.9420 & -1.9670 & 0.0808 & 0.0065 & 0.9540 & -1.9737 & 0.0846 & 0.0071 & 0.9460 \\
 & $\beta_1^R$ & 0.5000 & 0.4964 & 0.0703 & 0.0049 & 0.9500 & 0.4983 & 0.0704 & 0.0049 & 0.9620 & 0.5004 & 0.0719 & 0.0052 & 0.9480 \\
 & $\beta_2^R$ & -0.4000 & -0.3968 & 0.0716 & 0.0051 & 0.9500 & -0.4020 & 0.0727 & 0.0053 & 0.9420 & -0.4028 & 0.0745 & 0.0055 & 0.9380 \\
 & $\beta_3^R$ & 0.4000 & 0.4022 & 0.0703 & 0.0049 & 0.9500 & 0.4019 & 0.0711 & 0.0050 & 0.9540 & 0.4022 & 0.0734 & 0.0054 & 0.9320 \\
 & $\gamma_1^R$ & -3.0000 & -3.0180 & 0.4327 & 0.1871 & 0.9400 & -3.0339 & 0.4612 & 0.2135 & 0.9380 & -2.9648 & 0.4348 & 0.1899 & 0.9440 \\
 & $\log\tau$ & 3.4012 & 3.4081 & 0.0452 & 0.0021 & 0.9440 & 3.4083 & 0.0436 & 0.0019 & 0.9480 & 3.4115 & 0.0431 & 0.0020 & 0.9420 \\
\end{tabular}}}
\end{table}

\begin{table}[H]
\centering
\caption{\footnotesize Parameter estimates, SDs, MSEs and CPs for CCR-KvM model with three linear and one circular covariate.}
\label{tabK4}
\fbox{
\resizebox{0.9\linewidth}{!}{%
\begin{tabular}{llc cccc cccc cccc}
& & & \multicolumn{4}{c}{$\delta=0$} & \multicolumn{4}{c}{$\delta=5$} & \multicolumn{4}{c}{$\delta=-5$} \\
\cmidrule(lr){4-7} \cmidrule(lr){8-11} \cmidrule(lr){12-15}
Sample size & Parameter & True & Estimate & SD & MSE & CP & Estimate & SD & MSE & CP & Estimate & SD & MSE & CP \\
\midrule
\multirow{14}{*}{100} & $\mathcal{M}^C$ & 1.0000 & 0.9997 & 0.0421 & 0.0018 & 0.9340 & 1.0016 & 0.0595 & 0.0035 & 0.9140 & 1.0017 & 0.0281 & 0.0008 & 0.9420 \\
 & $\beta_1^C$ & 0.4000 & 0.4010 & 0.0261 & 0.0007 & 0.9280 & 0.4010 & 0.0345 & 0.0012 & 0.9440 & 0.4019 & 0.0186 & 0.0004 & 0.9340 \\
 & $\beta_2^C$ & -0.3500 & -0.3523 & 0.0253 & 0.0006 & 0.9420 & -0.3510 & 0.0350 & 0.0012 & 0.9440 & -0.3513 & 0.0171 & 0.0003 & 0.9280 \\
 & $\beta_3^C$ & 0.3000 & 0.3000 & 0.0231 & 0.0005 & 0.9500 & 0.3004 & 0.0318 & 0.0010 & 0.9500 & 0.3009 & 0.0162 & 0.0003 & 0.9440 \\
 & $\gamma_{11}^C$ & 0.4000 & 0.4007 & 0.0374 & 0.0014 & 0.9500 & 0.4015 & 0.0541 & 0.0029 & 0.9260 & 0.3992 & 0.0267 & 0.0007 & 0.9260 \\
 & $\gamma_{12}^C$ & -0.3000 & -0.3002 & 0.0346 & 0.0012 & 0.9220 & -0.3011 & 0.0450 & 0.0020 & 0.9360 & -0.3015 & 0.0221 & 0.0005 & 0.9280 \\
 & $\zeta$ & 2.3026 & 2.3571 & 0.3357 & 0.1155 & 0.9380 & 2.3888 & 0.3481 & 0.1283 & 0.9320 & 2.3199 & 0.3149 & 0.0992 & 0.9420 \\
 & $\delta$ & 0 & -0.1890 & 2.3607 & 5.5976 & 0.9280 & 5.1281 & 2.4043 & 5.7854 & 0.9300 & -5.4790 & 2.2958 & 5.4894 & 0.9460 \\
 & $\mathcal{M}^R$ & 3.9000 & 4.0834 & 0.6465 & 0.4507 & 0.9240 & 4.0754 & 0.6295 & 0.4263 & 0.9360 & 4.0882 & 0.6279 & 0.4289 & 0.9440 \\
 & $\beta_1^R$ & -0.5000 & -0.5428 & 0.4932 & 0.2446 & 0.9440 & -0.5305 & 0.4972 & 0.2477 & 0.9320 & -0.5313 & 0.4848 & 0.2355 & 0.9360 \\
 & $\beta_2^R$ & 0.4000 & 0.3922 & 0.4566 & 0.2081 & 0.9500 & 0.4524 & 0.4617 & 0.2155 & 0.9540 & 0.4191 & 0.4748 & 0.2253 & 0.9460 \\
 & $\beta_3^R$ & -0.4000 & -0.4036 & 0.4716 & 0.2220 & 0.9460 & -0.4158 & 0.4901 & 0.2400 & 0.9360 & -0.4174 & 0.4734 & 0.2240 & 0.9360 \\
 & $\gamma_1^R$ & 3.0000 & 3.1152 & 2.7585 & 7.6071 & 0.9580 & 3.1382 & 2.8788 & 8.2902 & 0.9420 & 3.2782 & 2.8903 & 8.4145 & 0.9540 \\
 & $\log a$ & 0.6931 & 0.7288 & 0.0816 & 0.0079 & 0.9140 & 0.7347 & 0.0857 & 0.0091 & 0.8980 & 0.7338 & 0.0847 & 0.0088 & 0.9200 \\
\midrule
\multirow{14}{*}{500} & $\mathcal{M}^C$ & 1.0000 & 1.0011 & 0.0186 & 0.0003 & 0.9420 & 0.9995 & 0.0241 & 0.0006 & 0.9620 & 0.9999 & 0.0128 & 0.0002 & 0.9360 \\
 & $\beta_1^C$ & 0.4000 & 0.3995 & 0.0106 & 0.0001 & 0.9640 & 0.4009 & 0.0142 & 0.0002 & 0.9620 & 0.4001 & 0.0078 & 0.0001 & 0.9520 \\
 & $\beta_2^C$ & -0.3500 & -0.3500 & 0.0101 & 0.0001 & 0.9560 & -0.3502 & 0.0141 & 0.0002 & 0.9520 & -0.3498 & 0.0075 & 0.0001 & 0.9380 \\
 & $\beta_3^C$ & 0.3000 & 0.2990 & 0.0104 & 0.0001 & 0.9340 & 0.3004 & 0.0144 & 0.0002 & 0.9460 & 0.2995 & 0.0069 & 0.0000 & 0.9580 \\
 & $\gamma_{11}^C$ & 0.4000 & 0.3985 & 0.0161 & 0.0003 & 0.9560 & 0.4021 & 0.0207 & 0.0004 & 0.9620 & 0.4003 & 0.0110 & 0.0001 & 0.9460 \\
 & $\gamma_{12}^C$ & -0.3000 & -0.3006 & 0.0139 & 0.0002 & 0.9540 & -0.3015 & 0.0197 & 0.0004 & 0.9520 & -0.3001 & 0.0093 & 0.0001 & 0.9700 \\
 & $\zeta$ & 2.3026 & 2.3161 & 0.1382 & 0.0193 & 0.9360 & 2.3277 & 0.1327 & 0.0182 & 0.9480 & 2.3105 & 0.1300 & 0.0169 & 0.9500 \\
 & $\delta$ & 0 & -0.0310 & 0.9562 & 0.9134 & 0.9500 & 5.0468 & 0.8809 & 0.7767 & 0.9440 & -5.0437 & 0.9259 & 0.8575 & 0.9640 \\
 & $\mathcal{M}^R$ & 3.9000 & 3.9394 & 0.2656 & 0.0720 & 0.9560 & 3.9426 & 0.2678 & 0.0734 & 0.9480 & 3.9317 & 0.2793 & 0.0788 & 0.9320 \\
 & $\beta_1^R$ & -0.5000 & -0.5119 & 0.1940 & 0.0377 & 0.9560 & -0.5127 & 0.2027 & 0.0412 & 0.9340 & -0.4938 & 0.2042 & 0.0417 & 0.9440 \\
 & $\beta_2^R$ & 0.4000 & 0.4050 & 0.2020 & 0.0407 & 0.9320 & 0.4047 & 0.1903 & 0.0362 & 0.9540 & 0.4029 & 0.1916 & 0.0367 & 0.9460 \\
 & $\beta_3^R$ & -0.4000 & -0.4011 & 0.2032 & 0.0412 & 0.9340 & -0.4004 & 0.1963 & 0.0385 & 0.9460 & -0.4137 & 0.1999 & 0.0401 & 0.9360 \\
 & $\gamma_1^R$ & 3.0000 & 3.0267 & 1.2525 & 1.5663 & 0.9460 & 3.0281 & 1.2323 & 1.5163 & 0.9480 & 3.1129 & 1.2345 & 1.5337 & 0.9460 \\
 & $\log a$ & 0.6931 & 0.7018 & 0.0351 & 0.0013 & 0.9540 & 0.7019 & 0.0365 & 0.0014 & 0.9440 & 0.7015 & 0.0375 & 0.0015 & 0.9380 \\
\midrule
\multirow{14}{*}{1000} & $\mathcal{M}^C$ & 1.0000 & 1.0007 & 0.0127 & 0.0002 & 0.9460 & 1.0018 & 0.0178 & 0.0003 & 0.9460 & 0.9996 & 0.0090 & 0.0001 & 0.9440 \\
 & $\beta_1^C$ & 0.4000 & 0.4005 & 0.0079 & 0.0001 & 0.9400 & 0.4000 & 0.0109 & 0.0001 & 0.9380 & 0.3998 & 0.0051 & 0.0000 & 0.9640 \\
 & $\beta_2^C$ & -0.3500 & -0.3502 & 0.0077 & 0.0001 & 0.9540 & -0.3499 & 0.0103 & 0.0001 & 0.9580 & -0.3502 & 0.0052 & 0.0000 & 0.9600 \\
 & $\beta_3^C$ & 0.3000 & 0.3001 & 0.0077 & 0.0001 & 0.9480 & 0.2998 & 0.0101 & 0.0001 & 0.9460 & 0.3001 & 0.0049 & 0.0000 & 0.9500 \\
 & $\gamma_{11}^C$ & 0.4000 & 0.3991 & 0.0119 & 0.0001 & 0.9300 & 0.3989 & 0.0157 & 0.0002 & 0.9440 & 0.4001 & 0.0081 & 0.0001 & 0.9420 \\
 & $\gamma_{12}^C$ & -0.3000 & -0.2999 & 0.0095 & 0.0001 & 0.9700 & -0.3005 & 0.0138 & 0.0002 & 0.9480 & -0.3003 & 0.0073 & 0.0001 & 0.9480 \\
 & $\zeta$ & 2.3026 & 2.3101 & 0.0977 & 0.0096 & 0.9260 & 2.3154 & 0.0932 & 0.0088 & 0.9440 & 2.3042 & 0.0950 & 0.0090 & 0.9540 \\
 & $\delta$ & 0 & 0.0019 & 0.6392 & 0.4078 & 0.9440 & 5.0466 & 0.6472 & 0.4202 & 0.9280 & -5.0620 & 0.6659 & 0.4464 & 0.9360 \\
 & $\mathcal{M}^R$ & 3.9000 & 3.9359 & 0.1948 & 0.0392 & 0.9340 & 3.9209 & 0.1922 & 0.0373 & 0.9440 & 3.9207 & 0.1813 & 0.0332 & 0.9540 \\
 & $\beta_1^R$ & -0.5000 & -0.5038 & 0.1433 & 0.0205 & 0.9500 & -0.5056 & 0.1481 & 0.0219 & 0.9220 & -0.5088 & 0.1401 & 0.0197 & 0.9600 \\
 & $\beta_2^R$ & 0.4000 & 0.3993 & 0.1433 & 0.0205 & 0.9420 & 0.4086 & 0.1325 & 0.0176 & 0.9540 & 0.3898 & 0.1363 & 0.0186 & 0.9580 \\
 & $\beta_3^R$ & -0.4000 & -0.4034 & 0.1422 & 0.0202 & 0.9520 & -0.4044 & 0.1403 & 0.0197 & 0.9380 & -0.3960 & 0.1441 & 0.0207 & 0.9380 \\
 & $\gamma_1^R$ & 3.0000 & 2.9825 & 0.8507 & 0.7226 & 0.9440 & 3.0158 & 0.8918 & 0.7939 & 0.9400 & 3.0850 & 0.8321 & 0.6982 & 0.9520 \\
 & $\log a$ & 0.6931 & 0.6998 & 0.0254 & 0.0007 & 0.9360 & 0.6977 & 0.0264 & 0.0007 & 0.9380 & 0.6976 & 0.0249 & 0.0006 & 0.9520 \\
\end{tabular}}}
\end{table}

\section{Application} \label{sec:application}
In this section, we demonstrate the practical utility of our proposed model using two wind direction datasets from South Africa and Brazil, respectively. We make use of the interval-valued circular mean error
(IN-CME) from \cite{mohan2026interval} given as 
\begin{equation*}
\operatorname{IN\text{-}CME}
(\widehat{\boldsymbol{\theta}},\boldsymbol{\theta})
=
\frac{1}{N}
\sum_{i=1}^{N}
\left[
d\left(\theta_i^U,\widehat{\theta}_i^U\right)
+
d\left(\theta_i^L,\widehat{\theta}_i^L\right)
\right],
\label{IN-CME}
\end{equation*}
where $d(\theta,\widehat{\theta})=1-\cos(\theta-\widehat{\theta}).
$

\begin{figure}[H]
\centering
\includegraphics[width=0.9\linewidth]{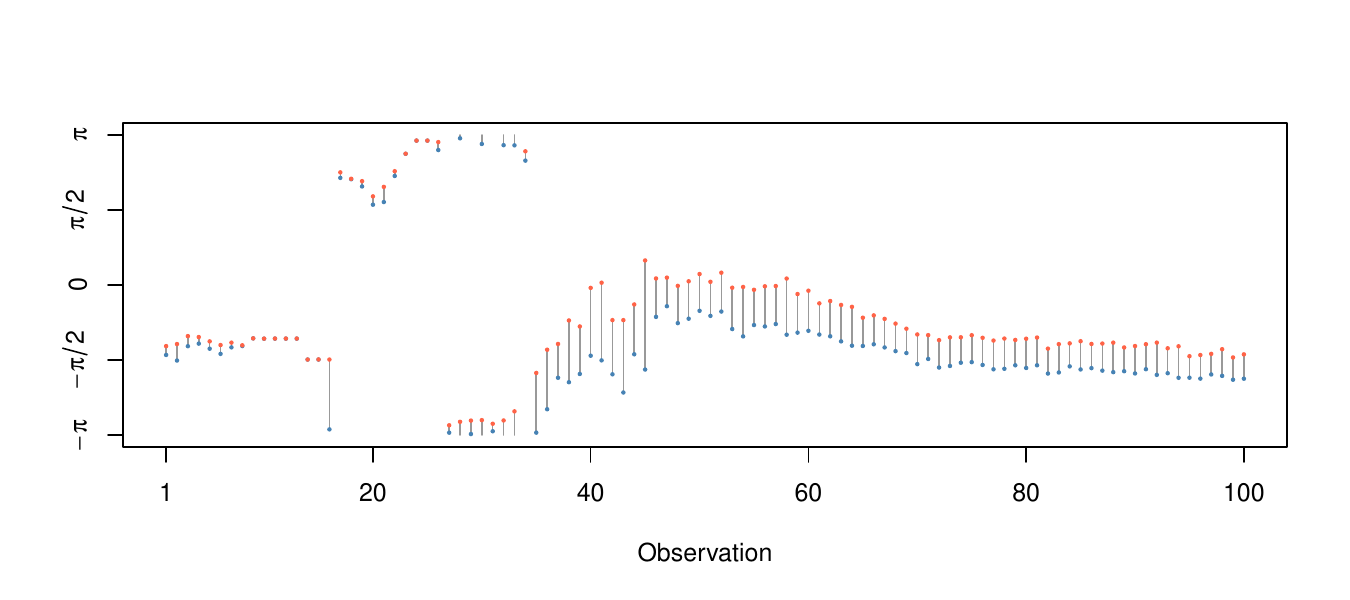}
\includegraphics[width=0.9\linewidth]{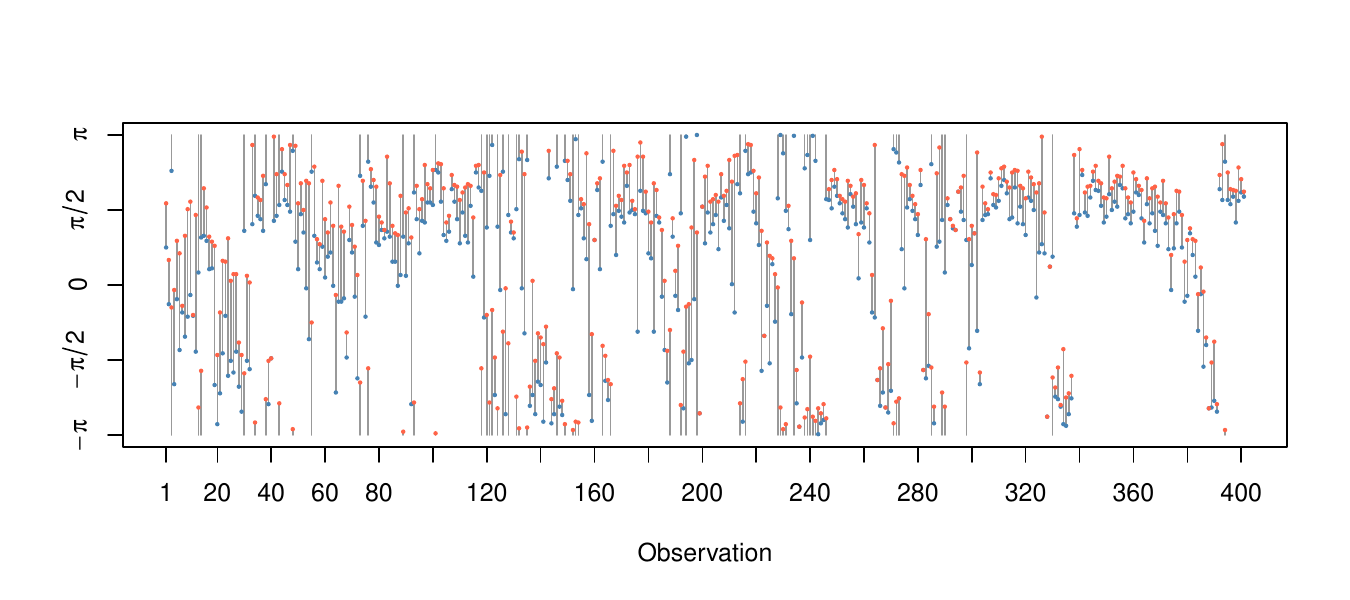}
\caption{\footnotesize Interval-valued wind direction observations for the Noupoort dataset (top)
and the Brazil dataset (bottom).}
\label{fig:apps}
\end{figure}

To test whether the centre and range of the intervals are dependent, we formulate
$H_0:\delta=0$ against $H_1:\delta\neq0$. Under $H_0$, the concentration of the centre does not
depend on the range, so that the centre and range are conditionally independent given the
covariates. The likelihood ratio statistic (LR) is
\begin{equation}\label{eq:LR}
LR=2\Big\{\ell(\hat{\boldsymbol{\psi}})-\ell(\tilde{\boldsymbol{\psi}})\Big\}
=2\Big\{\ell_{C\mid Q}(\hat{\boldsymbol{\psi}}_C)-\ell_{C\mid Q}(\tilde{\boldsymbol{\psi}}_C)\Big\},
\end{equation}
where $\hat{\boldsymbol{\psi}}$ is the likelihood assuming $\delta \neq 0$ and 
$\tilde{\boldsymbol{\psi}}$ is the parameter estimates under $H_0$. The LR statistic has a $\chi^2_1$ distribution and
$H_0$ is rejected at the $5\%$ level if LR $>3.841$.\\\\
The first dataset was recorded at a meteorological location, the Nouport wind farm in South
Africa, which records wind direction, wind speed, air temperature and relative humidity and 100
observations were used for this application. The response interval-valued variable is the
interval-valued wind direction measured at a height of 60\,m, and the covariates are the
interval-valued wind speed at the same height, air temperature and relative humidity, treated
as linear covariates, together with the time of day, treated as a circular covariate. The second dataset was obtained from the Brazil weather platform \footnote{\url{https://gee-community-catalog.org/projects/br_dwgd/}} and consists of 
meteorological observations recorded at a weather station in Brazil. There were 401 observations used in this application. The response observation is the interval-valued wind
direction, and the covariates are the interval-valued dew point temperature, relative humidity
and wind speed, treated as linear covariates, together with the time of day, treated as a
circular covariate. The circular and linear interval-valued covariates are represented using their corresponding centre and range representations. The plots of the observations from the circular response variables are provided in Figure \ref{fig:apps}.

In Tables~\ref{tab7} and~\ref{BRA}, the estimates, 95\% confidence intervals, IN-CME,  Bayesian Information Criterion (BIC), LR statistic and p-value for the test are provided for both datasets, respectively. The two regression models for which $\delta$ is estimated and $\delta=0$ are reported. The significant parameter estimates have been shown with asterisks.

For the Noupoort dataset, wind speed and the sine term of the time of day have a significant
effect on the mean direction of the intervals, whereas temperature, humidity and the cosine
term of the time of day are not significant. In the range equation, the ranges of wind speed
and humidity are significant for all three models, so that intervals with more variable wind
speed and humidity have wider wind direction intervals, and the range of the temperature is
significant for the CCR-BvM model. Wind speed therefore affects both the location and the width
of the intervals. The estimate of the dependence parameter is significant and negative,
indicating that the concentration of the centre depends on the range of the interval. The
dependent CCR-BvM model has the lowest BIC amongst the three models.

For the Noupoort dataset, all intervals span 20 minutes, so the range of the time-of-day
covariate is constant and is not included in the range equation.

For the Brazil dataset, dew point temperature, relative humidity and the cosine term of the time
of day have a significant effect on the mean direction of the intervals, whereas wind speed and
the sine term of the time of day are not significant. In the range equation, the ranges of the
time covariate and the wind speed are significant for all three models. The range of the dew
point temperature is significant for the CCR-BvM and CCR-KvM models, and the range of the
humidity for the CCR-BvM model, so that dew point temperature and humidity affect both the
location and the width of the intervals. As for the Noupoort dataset, the dependence between the
centre and the range is significant, but the estimate of the dependence parameter is positive.
Additionally, for the Brazil dataset, the dependent CCR-BvM model has the lowest BIC.

The two applications illustrate the flexibility of the proposed models. The centre and range
equations can depend on different covariates, and the dependence between the centre and the
range is significant at both sites but of opposite sign. In both applications, the beta
distribution gives the best description of the range.

\begin{table}[H]
\centering
\caption{\footnotesize Parameter estimates and 95\% confidence intervals for the CCR models fitted to the Noupoort dataset, with $\delta$ estimated and with $\delta=0$.}
\label{tab7}
\fbox{
{\scriptsize
\setlength{\tabcolsep}{3pt}
\begin{tabular}{llcccccc}
& & \multicolumn{3}{c}{Dependent} & \multicolumn{3}{c}{Independent} \\
\cmidrule(lr){3-5} \cmidrule(lr){6-8}
Model & Parameter & Estimate & CI lower & CI upper & Estimate & CI lower & CI upper \\
\midrule
\multirow{16}{*}{CCR-TWvM} & $\mathcal{M}^C$ & -1.5723$^{*}$ & -1.6862 & -1.4584 & -1.5931$^{*}$ & -1.7244 & -1.4619 \\
 & $\beta_1^C$ (wind speed) & 0.2181$^{*}$ & 0.1385 & 0.2977 & 0.1821$^{*}$ & 0.0939 & 0.2703 \\
 & $\beta_2^C$ (temperature) & 0.1699 & -0.0093 & 0.3492 & 0.0741 & -0.1798 & 0.3280 \\
 & $\beta_3^C$ (humidity) & -0.0088 & -0.1899 & 0.1723 & -0.0720 & -0.3023 & 0.1584 \\
 & $\gamma_{11}^C$ (time, cos) & -0.1472 & -0.3435 & 0.0490 & -0.0941 & -0.3598 & 0.1716 \\
 & $\gamma_{12}^C$ (time, sin) & -0.2601$^{*}$ & -0.3795 & -0.1408 & -0.3783$^{*}$ & -0.5465 & -0.2102 \\
 & $\zeta$ & 0.4240$^{*}$ & 0.0436 & 0.8045 & 1.1271$^{*}$ & 0.8861 & 1.3681 \\
 & $\delta$ & -11.3152$^{*}$ & -14.5732 & -8.0573 &--  & -- & -- \\
 & $\mathcal{M}^R$ & -4.0293$^{*}$ & -4.6550 & -3.4036 & -4.0293$^{*}$ & -4.6550 & -3.4036 \\
 & $\beta_1^R$ (wind speed) & 0.8247$^{*}$ & 0.4656 & 1.1839 & 0.8247$^{*}$ & 0.4656 & 1.1839 \\
 & $\beta_2^R$ (temperature) & 1.5415 & -1.7756 & 4.8586 & 1.5415 & -1.7756 & 4.8586 \\
 & $\beta_3^R$ (humidity) & 1.3516$^{*}$ & 0.1520 & 2.5513 & 1.3516$^{*}$ & 0.1520 & 2.5513 \\
 & $\log r$ & 0.0783 & -0.0885 & 0.2451 & 0.0783 & -0.0885 & 0.2451 \\
\cmidrule(lr){2-8}
 & IN-CME & \multicolumn{3}{c}{0.4089} & \multicolumn{3}{c}{0.3909} \\
 & BIC & \multicolumn{3}{c}{-99.5272} & \multicolumn{3}{c}{-69.0974} \\
 & LR ($p$-value) & \multicolumn{6}{c}{35.0349  ($<0.0001$)} \\
\midrule
\multirow{16}{*}{CCR-BvM} & $\mathcal{M}^C$ & -1.5723$^{*}$ & -1.6862 & -1.4584 & -1.5931$^{*}$ & -1.7244 & -1.4619 \\
 & $\beta_1^C$ (wind speed) & 0.2181$^{*}$ & 0.1385 & 0.2977 & 0.1821$^{*}$ & 0.0939 & 0.2703 \\
 & $\beta_2^C$ (temperature) & 0.1699 & -0.0093 & 0.3492 & 0.0741 & -0.1798 & 0.3280 \\
 & $\beta_3^C$ (humidity) & -0.0088 & -0.1899 & 0.1723 & -0.0720 & -0.3023 & 0.1584 \\
 & $\gamma_{11}^C$ (time, cos) & -0.1472 & -0.3435 & 0.0490 & -0.0941 & -0.3598 & 0.1716 \\
 & $\gamma_{12}^C$ (time, sin) & -0.2601$^{*}$ & -0.3795 & -0.1408 & -0.3783$^{*}$ & -0.5465 & -0.2102 \\
 & $\zeta$ & 0.4240$^{*}$ & 0.0436 & 0.8045 & 1.1271$^{*}$ & 0.8861 & 1.3681 \\
 & $\delta$ & -11.3152$^{*}$ & -14.5732 & -8.0573 & -- & -- & -- \\
 & $\mathcal{M}^R$ & -4.1764$^{*}$ & -4.5217 & -3.8312 & -4.1764$^{*}$ & -4.5217 & -3.8312 \\
 & $\beta_1^R$ (wind speed) & 1.0092$^{*}$ & 0.8055 & 1.2129 & 1.0092$^{*}$ & 0.8055 & 1.2129 \\
 & $\beta_2^R$ (temperature) & 1.3060$^{*}$ & 0.1256 & 2.4865 & 1.3060$^{*}$ & 0.1256 & 2.4865 \\
 & $\beta_3^R$ (humidity) & 1.3554$^{*}$ & 0.8729 & 1.8380 & 1.3554$^{*}$ & 0.8729 & 1.8380 \\
 & $\log\tau$ & 3.2977$^{*}$ & 3.0116 & 3.5839 & 3.2977$^{*}$ & 3.0116 & 3.5839 \\
\cmidrule(lr){2-8}
 & IN-CME & \multicolumn{3}{c}{0.4078} & \multicolumn{3}{c}{0.3899} \\
 & BIC & \multicolumn{3}{c}{-180.8793} & \multicolumn{3}{c}{-150.4496} \\
 & LR ($p$-value) & \multicolumn{6}{c}{35.0349  ($<0.0001$)} \\
\midrule
\multirow{16}{*}{CCR-KvM} & $\mathcal{M}^C$ & -1.5723$^{*}$ & -1.6862 & -1.4584 & -1.5931$^{*}$ & -1.7244 & -1.4619 \\
 & $\beta_1^C$ (wind speed) & 0.2181$^{*}$ & 0.1385 & 0.2977 & 0.1821$^{*}$ & 0.0939 & 0.2703 \\
 & $\beta_2^C$ (temperature) & 0.1699 & -0.0093 & 0.3492 & 0.0741 & -0.1798 & 0.3280 \\
 & $\beta_3^C$ (humidity) & -0.0088 & -0.1899 & 0.1723 & -0.0720 & -0.3023 & 0.1584 \\
 & $\gamma_{11}^C$ (time, cos) & -0.1472 & -0.3435 & 0.0490 & -0.0941 & -0.3598 & 0.1716 \\
 & $\gamma_{12}^C$ (time, sin) & -0.2601$^{*}$ & -0.3795 & -0.1408 & -0.3783$^{*}$ & -0.5465 & -0.2102 \\
 & $\zeta$ & 0.4240$^{*}$ & 0.0436 & 0.8045 & 1.1271$^{*}$ & 0.8861 & 1.3681 \\
 & $\delta$ & -11.3152$^{*}$ & -14.5732 & -8.0573 & --& -- & -- \\
 & $\mathcal{M}^R$ & 4.0623$^{*}$ & 3.2782 & 4.8464 & 4.0623$^{*}$ & 3.2782 & 4.8464 \\
 & $\beta_1^R$ (wind speed) & -0.8630$^{*}$ & -1.2298 & -0.4962 & -0.8630$^{*}$ & -1.2298 & -0.4962 \\
 & $\beta_2^R$ (temperature) & -1.3301 & -4.3169 & 1.6567 & -1.3301 & -4.3169 & 1.6567 \\
 & $\beta_3^R$ (humidity) & -1.4385$^{*}$ & -2.6146 & -0.2623 & -1.4385$^{*}$ & -2.6146 & -0.2623 \\
 & $\log a$ & 0.0168 & -0.1674 & 0.2009 & 0.0168 & -0.1674 & 0.2009 \\
\cmidrule(lr){2-8}
 & IN-CME & \multicolumn{3}{c}{0.4066} & \multicolumn{3}{c}{0.3887} \\
 & BIC & \multicolumn{3}{c}{-103.9238} & \multicolumn{3}{c}{-73.4940} \\
 & LR ($p$-value) & \multicolumn{6}{c}{35.0349 
 ($<0.0001$)} \\
\end{tabular}}}
\\[2pt]{\scriptsize $^{*}$Significant at the 5\% level.}
\end{table}

\begin{table}[H]
\centering
\caption{\footnotesize Parameter estimates and 95\% confidence intervals for the CCR models fitted to the Brazil weather dataset, with $\delta$ estimated and with $\delta=0$.}
\label{BRA}
\fbox{
{\scriptsize
\setlength{\tabcolsep}{3pt}
\begin{tabular}{llcccccc}
& & \multicolumn{3}{c}{Dependent} & \multicolumn{3}{c}{Independent} \\
\cmidrule(lr){3-5} \cmidrule(lr){6-8}
Model & Parameter & Estimate & CI lower & CI upper & Estimate & CI lower & CI upper \\
\midrule
\multirow{17}{*}{CCR-TWvM} & $\mathcal{M}^C$ & 1.8661$^{*}$ & 1.7001 & 2.0320 & 1.8981$^{*}$ & 1.7177 & 2.0784 \\
 & $\beta_1^C$ (dew point) & -0.1209$^{*}$ & -0.2182 & -0.0237 & -0.1461$^{*}$ & -0.2615 & -0.0307 \\
 & $\beta_2^C$ (humidity) & 0.2726$^{*}$ & 0.1364 & 0.4088 & 0.2623$^{*}$ & 0.1100 & 0.4146 \\
 & $\beta_3^C$ (wind speed) & 0.0576 & -0.0344 & 0.1495 & 0.0102 & -0.0927 & 0.1131 \\
 & $\gamma_{11}^C$ (time, cos) & 0.3785$^{*}$ & 0.2414 & 0.5157 & 0.3300$^{*}$ & 0.1803 & 0.4798 \\
 & $\gamma_{12}^C$ (time, sin) & 0.0076 & -0.1358 & 0.1510 & 0.0663 & -0.0972 & 0.2298 \\
 & $\zeta$ & 0.7086$^{*}$ & 0.4630 & 0.9542 & 0.0119 & -0.1624 & 0.1863 \\
 & $\delta$ & 4.1468$^{*}$ & 2.4010 & 5.8926 & -- & -- & -- \\
 & $\mathcal{M}^R$ & -3.5953$^{*}$ & -4.2543 & -2.9363 & -3.5953$^{*}$ & -4.2543 & -2.9363 \\
 & $\beta_1^R$ (dew point) & 0.2545 & -0.0291 & 0.5381 & 0.2545 & -0.0291 & 0.5381 \\
 & $\beta_2^R$ (humidity) & 0.1970 & -0.0220 & 0.4160 & 0.1970 & -0.0220 & 0.4160 \\
 & $\beta_3^R$ (wind speed) & 0.1484$^{*}$ & 0.0059 & 0.2909 & 0.1484$^{*}$ & 0.0059 & 0.2909 \\
 & $\gamma_1^R$ (time) & 6.3115$^{*}$ & 3.3368 & 9.2861 & 6.3115$^{*}$ & 3.3368 & 9.2861 \\
 & $\log r$ & 0.0855 & -0.0052 & 0.1762 & 0.0855 & -0.0052 & 0.1762 \\
\cmidrule(lr){2-8}
 & IN-CME & \multicolumn{3}{c}{1.1683} & \multicolumn{3}{c}{1.1667} \\
 & BIC & \multicolumn{3}{c}{654.2658} & \multicolumn{3}{c}{684.9992} \\
 & LR ($p$-value) & \multicolumn{6}{c}{44.9403 ($<0.0001$)} \\
\midrule
\multirow{17}{*}{CCR-BvM} & $\mathcal{M}^C$ & 1.8661$^{*}$ & 1.7001 & 2.0320 & 1.8981$^{*}$ & 1.7177 & 2.0784 \\
 & $\beta_1^C$ (dew point) & -0.1209$^{*}$ & -0.2182 & -0.0237 & -0.1461$^{*}$ & -0.2615 & -0.0307 \\
 & $\beta_2^C$ (humidity) & 0.2726$^{*}$ & 0.1364 & 0.4088 & 0.2623$^{*}$ & 0.1100 & 0.4146 \\
 & $\beta_3^C$ (wind speed) & 0.0576 & -0.0344 & 0.1495 & 0.0102 & -0.0927 & 0.1131 \\
 & $\gamma_{11}^C$ (time, cos) & 0.3785$^{*}$ & 0.2414 & 0.5157 & 0.3300$^{*}$ & 0.1803 & 0.4798 \\
 & $\gamma_{12}^C$ (time, sin) & 0.0076 & -0.1358 & 0.1510 & 0.0663 & -0.0972 & 0.2298 \\
 & $\zeta$ & 0.7086$^{*}$ & 0.4630 & 0.9542 & 0.0119 & -0.1624 & 0.1863 \\
 & $\delta$ & 4.1468$^{*}$ & 2.4010 & 5.8926 & -- & -- & -- \\
 & $\mathcal{M}^R$ & -3.3347$^{*}$ & -3.7677 & -2.9016 & -3.3347$^{*}$ & -3.7677 & -2.9016 \\
 & $\beta_1^R$ (dew point) & 0.2179$^{*}$ & 0.0511 & 0.3847 & 0.2179$^{*}$ & 0.0511 & 0.3847 \\
 & $\beta_2^R$ (humidity) & 0.1648$^{*}$ & 0.0164 & 0.3132 & 0.1648$^{*}$ & 0.0164 & 0.3132 \\
 & $\beta_3^R$ (wind speed) & 0.1253$^{*}$ & 0.0110 & 0.2397 & 0.1253$^{*}$ & 0.0110 & 0.2397 \\
 & $\gamma_1^R$ (time) & 6.4156$^{*}$ & 4.2087 & 8.6226 & 6.4156$^{*}$ & 4.2087 & 8.6226 \\
 & $\log\tau$ & 1.8813$^{*}$ & 1.7339 & 2.0287 & 1.8813$^{*}$ & 1.7339 & 2.0287 \\
\cmidrule(lr){2-8}
 & IN-CME & \multicolumn{3}{c}{1.1667} & \multicolumn{3}{c}{1.1650} \\
 & BIC & \multicolumn{3}{c}{597.1595} & \multicolumn{3}{c}{627.8929} \\
 & LR ($p$-value) & \multicolumn{6}{c}{44.9403 ($<0.0001$)} \\
\midrule
\multirow{17}{*}{CCR-KvM} & $\mathcal{M}^C$ & 1.8661$^{*}$ & 1.7001 & 2.0320 & 1.8981$^{*}$ & 1.7177 & 2.0784 \\
 & $\beta_1^C$ (dew point) & -0.1209$^{*}$ & -0.2182 & -0.0237 & -0.1461$^{*}$ & -0.2615 & -0.0307 \\
 & $\beta_2^C$ (humidity) & 0.2726$^{*}$ & 0.1364 & 0.4088 & 0.2623$^{*}$ & 0.1100 & 0.4146 \\
 & $\beta_3^C$ (wind speed) & 0.0576 & -0.0344 & 0.1495 & 0.0102 & -0.0927 & 0.1131 \\
 & $\gamma_{11}^C$ (time, cos) & 0.3785$^{*}$ & 0.2414 & 0.5157 & 0.3300$^{*}$ & 0.1803 & 0.4798 \\
 & $\gamma_{12}^C$ (time, sin) & 0.0076 & -0.1358 & 0.1510 & 0.0663 & -0.0972 & 0.2298 \\
 & $\zeta$ & 0.7086$^{*}$ & 0.4630 & 0.9542 & 0.0119 & -0.1624 & 0.1863 \\
 & $\delta$ & 4.1468$^{*}$ & 2.4010 & 5.8926 &  -- & -- & -- \\
 & $\mathcal{M}^R$ & 3.3819$^{*}$ & 2.7284 & 4.0353 & 3.3819$^{*}$ & 2.7284 & 4.0353 \\
 & $\beta_1^R$ (dew point) & -0.2543$^{*}$ & -0.5032 & -0.0055 & -0.2543$^{*}$ & -0.5032 & -0.0055 \\
 & $\beta_2^R$ (humidity) & -0.1692 & -0.3656 & 0.0272 & -0.1692 & -0.3656 & 0.0272 \\
 & $\beta_3^R$ (wind speed) & -0.1391$^{*}$ & -0.2693 & -0.0089 & -0.1391$^{*}$ & -0.2693 & -0.0089 \\
 & $\gamma_1^R$ (time) & -6.3238$^{*}$ & -9.3284 & -3.3193 & -6.3238$^{*}$ & -9.3284 & -3.3193 \\
 & $\log a$ & -0.0155 & -0.1161 & 0.0851 & -0.0155 & -0.1161 & 0.0851 \\
\cmidrule(lr){2-8}
 & IN-CME & \multicolumn{3}{c}{1.1649} & \multicolumn{3}{c}{1.1635} \\
 & BIC & \multicolumn{3}{c}{649.7083} & \multicolumn{3}{c}{680.4417} \\
 & LR ($p$-value) & \multicolumn{6}{c}{44.9403 ($<0.0001$)} \\
\end{tabular}}}
\\[2pt]{\scriptsize $^{*}$Significant at the 5\% level.}
\end{table}

\section{Concluding remarks}
    In this paper, we propose the circular centre-range (CCR) regression framework for interval-valued observations on the
    circle. This aligns with many real-world scenarios in which measurements are expressed as intervals. This approach respects the circular topology of the response, and hence each circular interval is represented by its centre and range. These representations are modelled jointly through a cylindrical distribution in a circular regression framework. 
    Linear and circular interval-valued
    covariates are explicitly incorporated by use of a link function. As with other link-based circular regression models, the proposed model is limited to
    covariate effects that rotate the mean direction within a bounded range. Parameters are estimated by a maximum likelihood approach. A simulation study demonstrated that the proposed estimation procedure performs well across various parameter settings and sample sizes, confirming the robustness of the proposed approach. An application to a real-world dataset from the Nouport wind farm in South Africa and a meteorological station in Brazil further validated the practical utility of the CCR  model. Future research could extend the framework to multivariate interval-valued circular
    responses.
     \section*{Acknowledgement}
     This work was based on research supported in part by the UCDP grant at the University of Pretoria, the National Research Foundation (NRF) of South Africa, Ref.: RA210106581084, grant No. 150170. The opinions expressed, and conclusions arrived at, are those of the authors and are not necessarily to be attributed to the NRF.\\\\
	\textbf{Data availability}: The South African data is available upon request and the Brazil data is freely available.\\\\
	\textbf{Declaration of Competing Interest}: The authors declare that they have no known competing financial interests or personal relationships that could have appeared to influence the work reported in this paper.
    
	\bibliographystyle{apalike}
	\bibliography{bibliography}
\end{document}